\documentclass{article}

\usepackage[utf8]{inputenc} 
\usepackage[T1]{fontenc}    
\usepackage[numbers,sort&compress,comma]{natbib}
\usepackage{balance}
\usepackage{mathptmx}
\usepackage{sectsty}
\usepackage{graphicx} 
\usepackage{lastpage}
\usepackage{float}
\usepackage{amsmath}
\usepackage{fancyhdr}
\usepackage{fnpos}
\usepackage[left=1.5cm, right=1.5cm, top=2.0cm, bottom=2.0cm]{geometry}
\usepackage[english]{babel}
\usepackage{array}
\usepackage{droidsans}
\usepackage{charter}
\usepackage{setspace}
\usepackage{xr-hyper}
\usepackage{hyperref}
\usepackage[noabbrev, capitalise]{cleveref}
\usepackage[compact]{titlesec}
\usepackage{amssymb}
\usepackage{authblk}
\usepackage{pdfpages}
\usepackage{caption}
\usepackage{multirow}
\usepackage{tabularray}
\usepackage{longtable}
\usepackage{ltablex}
\usepackage{boldline}
\usepackage{bm}
\usepackage{subfig}
\usepackage{soul}

\title{Pulsed laser ablation in liquid of sp-carbon chains: status and recent advances}

\author[1]{Pietro~Marabotti}
\author[1]{Sonia~Peggiani}
\author[1]{Alessandro~Vidale}
\author[1]{Carlo~Spartaco~Casari}
\affil[1]{Department of Energy, Politecnico di Milano, Via Ponzio 34/3, 20133, Milano, Italy}

\begin{document}
\captionsetup[figure]{labelfont=bf,labelformat=default,labelsep=period,name={Figure}}
\captionsetup[table]{labelfont=bf,labelformat=default,labelsep=period,name={Table}}
\date{}
\maketitle

\section*{Abstract}
\smallskip
This review provides a discussion of the current state of research on sp-carbon chains synthesized by pulsed laser ablation in liquid. In recent years, pulsed laser ablation in liquid (PLAL) has been widely employed for polyynes synthesis thanks to its flexibility with varying laser parameters, solvents, and targets. This allows the control of sp-carbon chains properties as yield, length, termination and stability. Although many reviews related to PLAL have been published, a comprehensive work reporting the current status and advances related to the synthesis of sp-carbon chains by PLAL is still missing. Here we first review the principle of PLAL and the mechanisms of formation of sp-carbon chains. Then we discuss the role of laser fluence (i.e. energy density), solvent, and target for sp-carbon chains synthesis. Lastly, we report the progress related to the prolonged stability of sp-carbon chains by PLAL encapsulated in polymeric matrices. This review will be a helpful guide for researchers interested in synthesizing sp-carbon chains by PLAL.

\section{Introduction}
The ability of carbon to produce a wide variety of nanostructures with different dimensionality and peculiar properties is demonstrated by the last thirty years of intense research on fullerenes (0-dimensional), nanotubes (quasi 1-dimensional), and graphene (2-dimensional). Besides these widely investigated systems, research has also looked for other carbon structures such as linear carbon and carbyne. Carbyne is the ideal linear carbon structure (1-dimensional) made by an atomic chain of sp-hybridized carbon atoms, representing the lacking carbon allotrope after graphite and diamond \cite{casari2016, casari2018, hirsch2010,banhart2020elemental}. 
While carbyne represents an ideal and infinite system, experimentally available systems are finite in length and endcapped by terminating atoms or functional groups. The longest sp-carbon chain ever reported is composed of 6000 atoms and is encapsulated in a carbon nanotube (the so-called confined carbyne)\cite{shi2016}. Isolated chains up to 48 sp-carbon atoms can be synthesized by chemical methods employing large terminating groups as stabilizing systems against crosslinking or even by polyyne rotaxanes \cite{gao2020loss,movsisyan2012synthesis,mov2016}.
The simplest finite sp-carbon chains are terminated by hydrogen atoms and can be prepared by physical techniques such as submerged arc discharge in liquid (SADL) and pulsed laser ablation in liquid (PLAL). Other physical techniques showing the synthesis of sp-carbon chains and sp-sp$^2$ carbon moieties are cluster sources based on laser ablation or arc discharge in gas. All these techniques share a common approach based on the production of a carbon plasma that is confined by a gas or a liquid to reach the out-of-equilibrium conditions needed for the formation of sp carbon chains (i.e. high temperature and pressure gradients).
Physical techniques in liquid are very efficient in the production of isolated carbon chains mainly in the form of polyynes (i.e., single-triple alternating bond structure), while only rarely has been reported the physical synthesis of cumulenes (i.e., all-double bond structure) \cite{cataldo1995photopolymerization, cataldo_CS2}.
PLAL is the most widely employed technique among the physical methods in liquid and shows advantages in the fine control of the ablation parameters and the low amount of carbonaceous byproducts present in the solution. PLAL allows one to employ non-conducting targets, non-solid targets such as suspensions, and even no target at all by directly ablating the solvent. Thanks to this versatility the research on the synthesis of polyynes by PLAL has seen advances in the last 15 years in terms of control of the produced polyynes, maximum length achieved, and understanding of the formation mechanisms by \textit{in situ} characterization. Moreover, PLAL has been shown to produce polyynes in a polymeric solution for the realization of nanocomposites \cite{peggianiPVA}.
Here we review the status and progress made in the synthesis of polyynes by PLAL with a focus on the role of the main process parameters in the polyyne formation. To this aim, the work is organized in a first section devoted to the discussion of the principles of PLAL and the mechanisms of formation of polyynes by PLAL. Three sections focus on the role of the main parameters affecting the polyyne formation, namely the laser fluence (\textit{i.e.} energy density), the solvent, and the target material. The last section reports the advances in the development of polyynes embedded in polymeric materials to form nanocomposites showing prolonged stability. Based on our work and the literature, the discussion aims at reporting a comprehensive review of the state of the art underlining the crucial aspects to be considered when synthesizing polyynes by PLAL.

\section{Synthesis of polyynes by pulsed laser ablation in liquid: principles and formation mechanisms} \label{formationmec_sec}
In this section, we will summarize the main phenomena that characterize pulsed laser ablation in liquid (PLAL), mainly focusing on the case of \textit{ns}-pulse lasers that were largely used in the production of sp-carbon chains. We will briefly introduce the main differences between ultrashort ($\approx 1 \ fs - 100 \ ps$) and short ($\approx 100 \ ps - 1 \ ns$) laser pulses. Moreover, we will illustrate the proposed mechanisms of formation of sp-carbon chains and the influence of the process parameters on the production, including a systematic study on PLAL parameters with acetonitrile. Finally, different \textit{in situ} approaches will be described to directly investigate the formation of sp-carbon chains by PLAL.

\subsection{Fundamental process occuring during PLAL}
\begin{figure}[t]
  \centering
  \includegraphics[width=\linewidth, keepaspectratio]{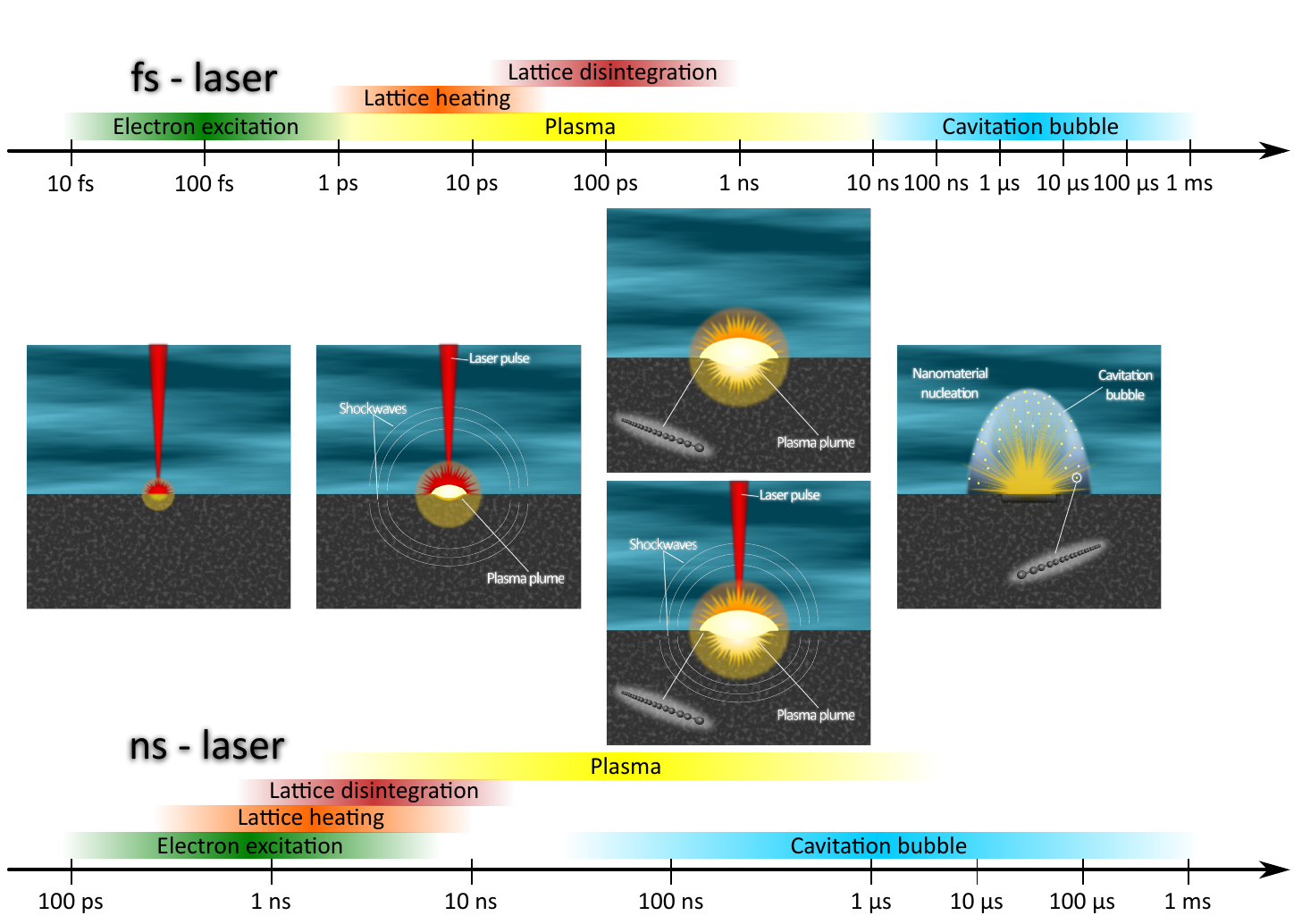}
  \caption{Scheme of the time evolution of the different phases of PLAL by \textit{fs} (upper timescale) or \textit{ns} (lower timescale) lasers. Formation of polyynes is assumed to happen during the plasma lifetime, though it has not been yet demonstrated. See the text for more details.}
  \label{fig:ablation_process}
\end{figure}

Before hitting the target, the laser beam passes through a liquid layer that could extend from a few millimeters to several centimeters in height (see \cref{fig:ablation_process}). The effect of the liquid layer on the laser beam should be considered when calculating the deposited laser fluence onto the target, to analyze possible defocusing effects and the optical absorption of the liquid itself. For example, it was found that the production of noble metal nanoparticles is influenced by the height of the solvent above the bulk target \cite{menendez2011transfer,nguyen2019impact,jiang2011promoting}. Moreover, the effect of the liquid meniscus on the laser spot dimension, and thus on the laser fluence, may be non-negligible, as discussed in Refs. \cite{menendez2011,marabotti2021situ}. In general, a good solvent (or solution) for PLAL experiments should be transparent at the selected laser wavelength to maximize the fraction of laser energy that reaches the target or, at most, the absorbed energy must be negligible compared to that reaching the target surface. Usually, this condition is fulfilled in PLAL experiments, even if some exceptions should be considered, for example, the presence of absorbing solute molecules, powders, or some nanomaterials (e.g. nanoparticles) that can significantly absorb the laser beam. The solvent employed in PLAL ranges from water to organic solvents, but also more unusual liquids (\textit{e.g.} ionic liquids or sol-gels) have been used. The laser fluence is usually limited to specific values to avoid optical breakdown or other nonlinear effects within the liquid \cite{berthe1999wavelength,nichols2006laser,kabashin2003synthesis,hoppius2019optimization}, even if some experiments may require these conditions. 

When the laser pulse hits the target, in general, the electrons of the target absorb the optical energy and are heavily heated up. Moreover, several nonlinear optical processes may occur, usually more pronounced with ultrashort pulses, such as multiphoton absorption and direct photoionization \cite{momma1996,kaiser2000microscopic,gruzdev2007photoionization,temnov2006multiphoton,hashida2009non}. Subsequently, the large amount of energy absorbed by the target induces the detachment of materials from the target by thermal and coulombic processes, within a volume approximately equal to the spot size \cite{momma1996,perez2003molecular,perez2008numerical}. The recoil pressure of the ablated material generates two shockwaves traveling both into the target and into the liquid (see \cref{fig:ablation_process}), at a supersonic speed ($> 10^3 m/s$) \cite{fabbro1990physical,sano1997residual,berthe1997shock}.
\\These highly ionized and energetic species can be considered as a strong out-of-equilibrium plasma. Plasma temperature, pressure, and density are approximately $10^3$ K, $10^{10}$ Pa, and $10^{22\div23}$ cm$^{-3}$, respectively \cite{amans2009nanodiamond,sajti2010gram,dellaglio2015mechanisms,fabbro1990physical,sakka2000laser,sakka2001laser}. If the laser pulse is comparable to the plasma lifetime (tens or hundreds of \textit{ns}, as reported in \cref{fig:ablation_process}), then the plasma plume can absorb the laser pulse - a phenomenon called plasma shielding - extending its duration and preventing further absorption of the laser beam by the target \cite{momma1996,yoo2000evidence,lu2002delayed,sakka2009spectral}. The plasma plume expands after the first 1-10 ns (see \cref{fig:ablation_process}) when it starts to be quenched by the surrounding solvent. Indeed, the presence of a liquid layer strongly confines the plasma plume compared to ablation experiments performed in vacuum or low-pressure gas atmosphere, reducing the area of target ablation to approximately the laser spot and increasing the temperature and pressure inside the plasma \cite{momma1996,sakka2000laser,kang2008laser,kumar2010synthesis}. At the plasma-liquid interface, the high temperatures cause the degradation, ionization, and pyrolisis of the solvent molecules \cite{sakka2000laser,sakka2002,peggianisolventi,amendola2009laser}. During this process, highly reactive species (e.g. radicals, fragments of molecules, and ions) can be generated and can react with the ablated target species \cite{boyer2012modeling,peggianisolventi,amendola2005synthesis,amendola2007free,amendola2009laser,begildayeva2021production}. It is reasonable to assume that these processes occur already during the plasma generation, thus from the picosecond time scale on, even if direct proofs are still missing.
\\As the plasma cools down, its thermal energy is transferred to the surrounding liquid that forms a hot vapor region, called cavitation bubble, approximately 1 $\mu s$ after the laser pulse (see \cref{fig:ablation_process}). The bubble travels with supersonic speed in the liquid, expanding and being cooled by the liquid. The formation and the collapse of the bubble are accompanied by the generation of shockwaves that propagate both toward the liquid and the target \cite{degiacomo2011laser,tanabe2015bubble,debonis2013dynamics,nguyen2013laser,tsuji2007nanosecond,tsuji2008preparation}. Although it is not clear if nanomaterials are formed during the plasma stage or within the cavitation bubble, it is reasonable to assume that the products of the ablation travel in the front of the cavitation bubble or inside it \cite{shchukin2006sonochemical,itina2011nanoparticle,nichols2006laser,nichols2006laser_2,tsuji2004npformation,tsuji2008preparation}. Finally, the products of the ablation process are rapidly spread over the surrounding liquid traveling with the fast shockwaves ($> 10^3 m/s$). Considering the repetition rates of the lasers employed in PLAL experiments, which usually are 1-100 Hz for \textit{ns}-lasers and 1-10 kHz for \textit{fs}-lasers, the nanomaterials are already far away from the ablation site before the laser hits again and thus can be considered as solutes for the next incoming laser pulse.
\\Depending on the pulse duration ($\tau_{pulse}$), some differences can exist in the processes occurring during PLAL. In principle, electronic and lattice heatings by the laser pulse are temporally separated occurring at a time scale of ($\approx 10^{-14 \div -12}$ s) and of ($\approx 10^{-12 \div -10}$ s), respectively (see \cref{fig:ablation_process}). This is true when \textit{fs}-lasers are employed when $\tau_{pulse}$ $\leq$ ps. In addition, nonlinear effects such as direct solvent dissociation, self-focusing, white light generation, and filamentation are far more probable in this case than in \textit{ns}-lasers \cite{besner2006,hoppius2019optimization,coua_2007}. When $\tau_{pulse} \geq ns$, instead, the absorption, heating, and energy transfer from the electrons to the lattice occur simultaneously (see \cref{fig:ablation_process}) and plasma plume and the laser pulse overlap. 
\\In the following, we will focus on \textit{ns}-PLAL since most of the works dealing with the physical synthesis of sp-carbon chains employ this kind of lasers.

\subsection{Mechanisms of formation of polyynes by PLAL}
Despite the increasing number of works on the investigation of sp-carbon chain properties and synthesis, a complete and fully comprehended description of the formation of polyynes during PLAL is still missing. One of the possible reasons can lie in the fact that the fast time scales of the ablation processes do not allow simple observation of the chemicophysical reactions that bring to the formation of sp-carbon chains. Moreover, the ablation environment is an intricate system to be modeled and the current numerical simulations cannot solve simultaneously the growth of nanomaterials, the evolution of a strongly confined plasma plume, and the fast chemical reactions involving both the solution and the target material. In general, it is widely acknowledged that sp-carbon chains are synthesized during the plasma plume phase of the ablation since these strong out-of-equilibrium conditions favor their formation (see \cref{fig:ablation_process}) \cite{tsuji2002,casari2016,cataldo_book}.
\begin{figure}
    \centering
    \includegraphics[width=\linewidth, keepaspectratio]{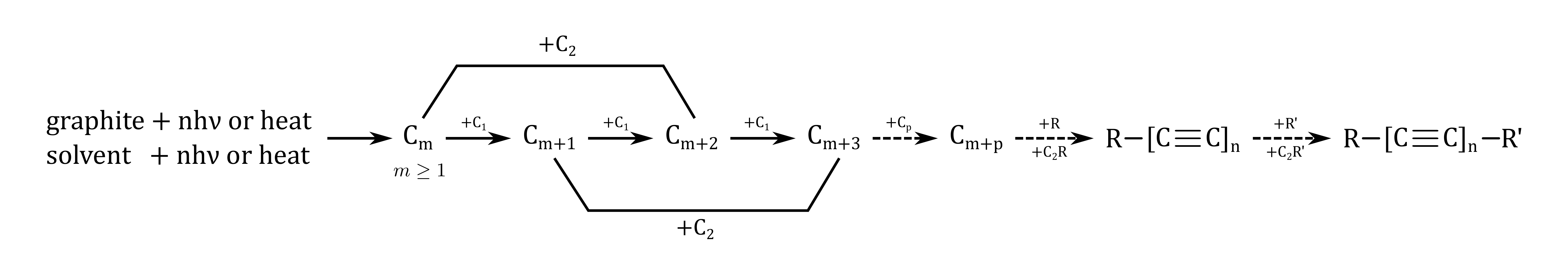}
    \caption{Scheme of the formation of polyynes during PLAL process. The polymerization reaction could occur \textit{via} carbon atoms (C$_1$), C$_2$ or larger (C$_p$) radicals. The termination process can involve different terminations (R and R').}
    \label{fig:formation_polyyne}
\end{figure}
\\In this framework, few models have been proposed over the years that try to illustrate the formation of sp-carbon chains during PLAL. In particular, these models refer to the formation of polyynes since none has detected cumulenes synthesized by PLAL so far, to the best of our knowledge. The most acknowledged model was advanced by Tsuji \textit{et al.}  \cite{tsuji2002} and is focused on \textit{ns}-PLAL. This model suggests that hydrogen-capped polyynes follow a step-wise growth by the addition of C$_1$ atoms or larger radical or fragments (from C$_2$ on), as shown in \cref{fig:formation_polyyne}. This elongation process, called polymerization, competes with the hydrogenation process that instead tends to terminate the chain with a hydrogen atom\cite{tsuji2002}. Moreover, in the case of a carbon-based target or powder, both the target and the solvent provide carbon atoms or radicals for the polymerization mechanism \cite{peggianisolventi}. Indeed, the high temperatures reached by the plasma plume cause the atomization of the several atomic layers of the target surface and induces the degradation of the solvent promoting the release of carbon atoms and/or C$_n$ radicals \cite{tsuji2002,tsuji2003,compagnini2008,peggianisolventi}. Concerning the hydrogenation process and assuming a target only made of carbon (e.g. graphite), the hydrogen atoms or ethynyl radicals C$_2$H (see \cref{fig:formation_polyyne}) can only come from the solvent \cite{tsuji2002,tsuji2003,peggianisolventi}. The hydrogenation process can also involve atoms different from hydrogen and in this case we talk about the \textit{termination} process, like CN, CH$_3$, or more complex terminations \cite{tsuji2002, peggianisolventi, forte2013, russo2014, wada2012, waka_2012, rama2015} as will be extensively described in Section \ref{solvent_sec}. In this sense, we performed an ablation of a graphite target in water where methyl-terminations were found even if no methyl radical can be directly produced by the solvent \cite{peggianisolventi}. Thus, we proposed that the formation of the -CH$_3$ termination happened by the addition of three H atoms to the terminating part of a growing chain. On the contrary, despite the use of aromatic compounds as solvents, like benzene and toluene\cite{tsuji2002}, none has demonstrated the formation of phenyl-capped polyynes or similar compounds. The lack of bulkier terminations in polyynes produced by PLAL is a consequence of the peculiar conditions of the plasma plume and its surroundings. Indeed, solvent molecules can be atomized or broken into several small fragments, thus if these building blocks reorganize themselves in aromatic rings, probably they tend to produce other structures (\textit{e.g.} sp$^2$ molecules) than terminate sp-carbon chains.
\\The degree of polymerization and termination of sp-carbon chains are strongly influenced by several ablation parameters: the pulse duration, wavelength, and pulse energy \cite{matsutani2011_CC, peggianisolventi, tsuji2005formation, tsuji2002, park2013}. Furthermore, the ablation solvent plays a critical role since it determines the liquid environment in which sp-carbon chains grow, varying the class of the possible terminations, the confinement and the thermal exchange rate with the plasma plume, and the stability of sp-carbon chains \cite{peggianisolventi,matsutani2011_CC, taguchi2015}. The radical production of the solvent is also an important aspect since the proposed mechanism for the growth of sp-carbon chains is based on radical reactions. The effect of the solvent will be discussed in more detail in Section \ref{solvent_sec}.
\\Regarding the effect of the pulse duration on the synthesis of polyynes by PLAL, \textit{ns} and \textit{fs} pulses were mostly employed. Despite the different dynamics of the overall ablation process (as shown in \cref{fig:ablation_process}), there are still no shreds of evidence that indicate different mechanisms of formation of sp-carbon chains by employing \textit{ns} or \textit{fs} lasers. Indeed, the growing mechanism of polyynes by \textit{fs}-laser ablation proposed by several authors is basically similar to the one of \textit{ns}-PLAL. The power densities reached in the filament produced by femtosecond lasers in the solvents are so high that the solvent undergoes direct dissociation in atomic, molecular, and parent ions \cite{hu2008,zaidi2010,zaidi2019femtosecond,sato2010,weso2011,zaidi2010}. In this way, short polyynes or cumulenes such as HC$_2$H, H$_2$C$_3$H$_2$, and HC$_4$ may be formed in a single-step process. Long polyynes, instead, cannot be formed directly from parent ions of the solvent and thus they must require several secondary reactions \cite{hu2008,zaidi2010,zaidi2019femtosecond,sato2010,weso2011,zaidi2010}. In particular, Zaidi \textit{et al.} \cite{zaidi2019femtosecond} proposed that long polyynes can be formed by reacting with shorter cumulenes or polyynes already formed with C$^+$, C$_2^+$ or C$_3^+$ ions, elongating the chains of several units. This last mechanism is quite similar to the radical growth experienced by polyynes in \textit{ns}-PLAL, except that here the building blocks are ionized species.
\\The effect of the laser wavelength on the synthesis of polyynes was analyzed by Matsutani \textit{et al.} \cite{matsutani2011_CC}, Park \textit{et al.} \cite{park2013}, and Tsuji \textit{et al.} \cite{tsuji2005formation}. The first two works employed solid targets and different solvents, methanol, and \textit{n}-hexane in Ref. \cite{matsutani2011_CC} and water in Ref. \cite{park2013}. They found that the production yield of polyynes and the maximum chain lengths grow with increasing wavelength. In this geometry, short-wavelength laser beams are more probably absorbed or scattered by impurities or byproducts of the ablation, thus diminishing the fraction of laser energy that reaches the target \cite{matsutani2011_CC}. Moreover, long polyynes (\textit{e.g.} from HC$_{20}$H on) start absorbing UV lasers and can be fragmented by the energetic laser beam \cite{matsutani2011_CC,marabotti2021situ}. On the contrary, based on the results of Tsuji \textit{et al.} \cite{tsuji2005formation}, if the target is a suspension of carbon-based nanoparticles, the formation yield of polyynes grows as the wavelength decreases regardless of the solvent. In these cases, we can assume that micro-plasma regions are formed in the surroundings ($\approx 100 \ \mu m$) of each carbon particle. As a consequence, due to the low concentration of carbon and terminating radicals, only small polyynes are produced in a lower quantity \cite{matsutani2011_CC,tsuji2005formation}. Moreover, shorter wavelengths have a larger probability to be absorbed by the carbon powder improving the yield in the case of graphite or fullerene powders \cite{matsutani2011_CC,tsuji2005formation}.

\subsection{Investigation of the formation of polyynes \textit{via} PLAL by \textit{in situ} techniques}
During the last years, the use of \textit{in situ} characterization techniques has largely expanded granting a direct investigation of several phenomena and processes in a wide variety of fields. Regarding the synthesis of nanomaterials by PLAL, several techniques have been employed to analyze the different aspects of PLAL depending on the objective. However, the time scales involved, from a few \textit{fs} to tens of \textit{ms}, complicate the analyses and reduce the number of characterization methods that can be exploited. Time-resolved shadowgraphy has been used to investigate the shape of the plasma plume and the cavitation bubble and to observe the shockwaves produced during the ablation process \cite{dellaglio2015mechanisms}. Optical emission spectroscopy, instead, is mainly exploited to characterize the species contained inside the plasma plume \cite{dellaglio2015mechanisms,sakka2000laser}. Other techniques can imply the use of x-rays, UV, or visible light to investigate the ongoing synthesis or specific properties of the growing nanomaterials.
\\ Only a few examples of \textit{in situ} analyses can be found in the literature regarding the synthesis of sp-carbon chains by PLAL. Hu \textit{et al.} \cite{hu2008}, Zaidi \textit{et al.} \cite{zaidi2010,zaidi2019femtosecond}, Sato \textit{et al.} \cite{sato2010}, Wesolowski \textit{et al.} \cite{weso2011} investigated the synthesis of polyynes by direct irradiation of solvents with femtosecond laser pulses. In these papers, they employed time of flight mass spectroscopy as the detection technique that, in this specific case, can be considered equivalent to an \textit{in situ} method. In this way, it is possible to investigate the products of the dissociation of the solvent or gas phase in which the ablation occurs and better understand the processes that lead to the formation of polyynes. Though they did not observe directly the growth of sp-carbon chains, these studies helped to understand the growing mechanism of polyynes in the case of \textit{fs}-PLAL without the use of a solid or powder target.
\begin{figure}[!t]
    \centering
    \includegraphics[height=13cm, keepaspectratio]{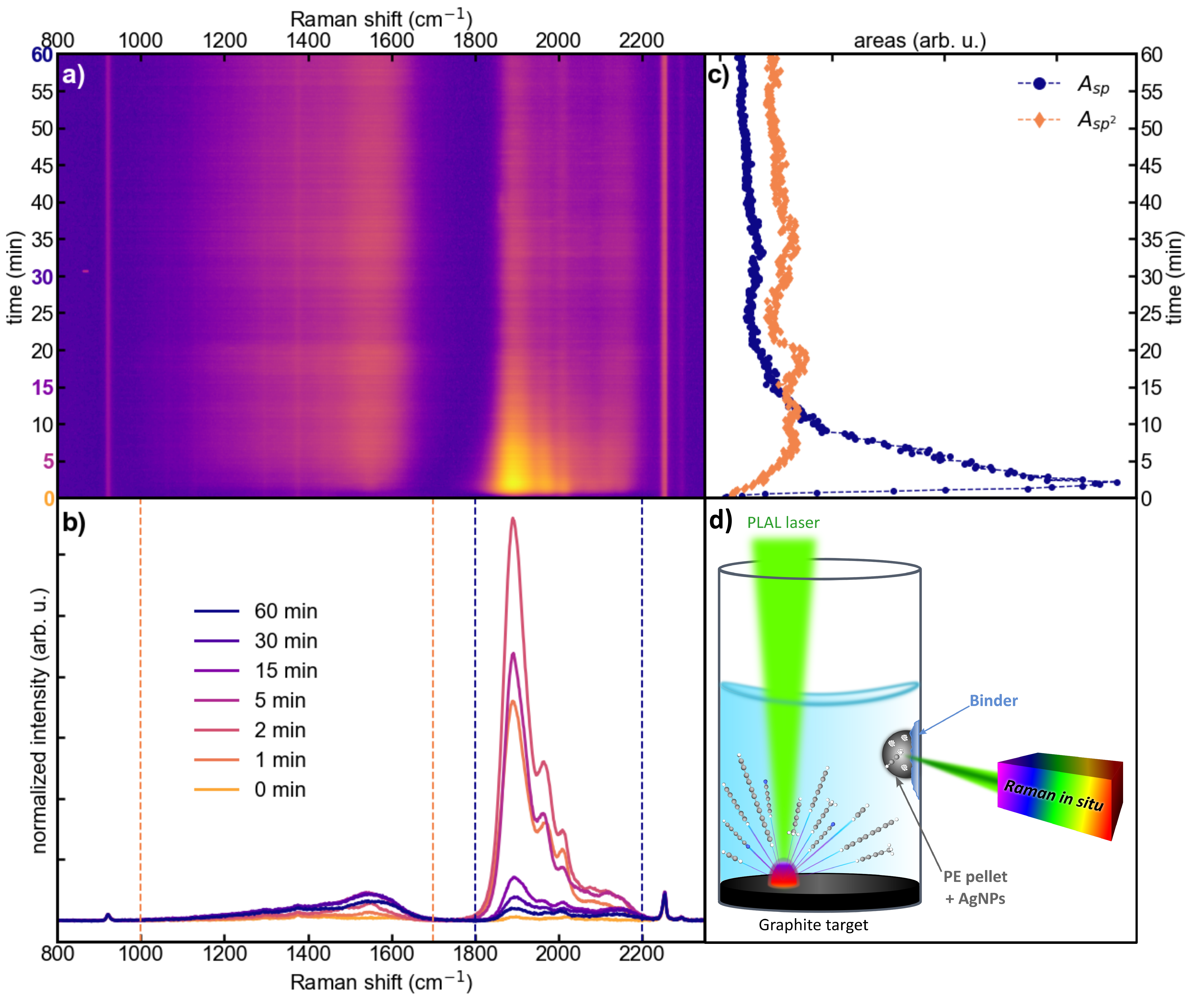}
    \caption{a) 2D plot of the evolution of the SERS signal during ablation in acetonitrile. b) SERS spectra of the species attached to the SERS substrate at fixed time intervals. Dashed colored vertical lines bound the sp$^2$ (1000-1700 cm$^{-1}$) and sp (1800-2200 cm$^{-1}$) spectral regions. c) Integrated areas of the sp$^2$ (A$_{sp}^2$ , 1000-1700 cm$^{-1}$) and sp (A$_{sp}$, 1800-2200 cm$^{-1}$) regions as a function of time. d) Scheme of \textit{in situ} surface-enhance Raman scattering setup. A polyethylene pellet functionalized with silver nanoparticles is attached to the glass vial with a polyvinyl alcohol layer as binder. The Raman laser beam is focused on the pellet surface from outside the glass vial. The ablation laser (PLAL laser) is directed on the graphite target far away from the pellet \cite{marabotti2021situ}}
    \label{fig:insitu_sers}
\end{figure}
\\Taguchi \textit{et al.} focused on the analysis of the plasma plume by optical emission spectroscopy generated by \textit{ns} and \textit{fs} ablation of hydrocarbon gas flow \cite{taguchi2017}. Again, by employing this characterization technique, it is possible to achieve information about the species in the plasma plume that are the building blocks for the formation of polyynes. The authors observed the characteristic C$_2$ Swan band in the emission spectra which is responsible for the formation of carbon clusters and also polyynes. They also have the hints of the production of a considerable amount of C$_2$H even if they did not have direct evidence from emission spectra due to the superposition of other strong signals. In conclusion, \textit{in situ} technique helped the authors to better understand the efficiency of production of long polyynes by changing the hydrocarbon gas.
\\In this framework, we studied the polyynes formation in solvents with nanosecond laser employing surface-enhanced Raman spectroscopy in two different works \cite{tesiPeggiani,marabotti2021situ}. In the first work, we developed a surface-enhanced Raman scattering substrate to investigate the formation of polyynes during PLAL in acetonitrile (see \cref{fig:insitu_sers}) \cite{marabotti2021situ}. We exploited the Raman signal enhancement given by the interaction of metal nanoparticles on the substrate with polyynes in the ablation solvent to acquire SERS spectra during 60 min of ablation in acetonitrile (see \cref{fig:insitu_sers}a). In such a way, we observed the signal of polyynes but also of the sp$^2$-like byproducts that we used to evaluate the efficiency of the synthesis and the ongoing degradation phenomena (see \cref{fig:insitu_sers}b and c). One of the advantages of this technique consists in the improved sensitivity that allowed us to record information about the synthesis even when the concentration of polyynes is too low for conventional Raman, in particular at the beginning of the ablation and in the case of longer species. Even if the time resolution of this technique cannot achieve the timescales of the formation of polyynes, our \textit{in situ} technique is the first one to achieve a direct visualization of the growth of polyynes during the ablation. Moreover, we found that degradation processes, as crosslinking reactions, start almost synchronously with the formation of polyynes by PLAL and they depend on the concentration of polyynes and byproducts, as well as on the length and termination of the chains.

\begin{figure}[!ht] 
	\centering
	\subfloat[][]{\includegraphics[scale=0.13]{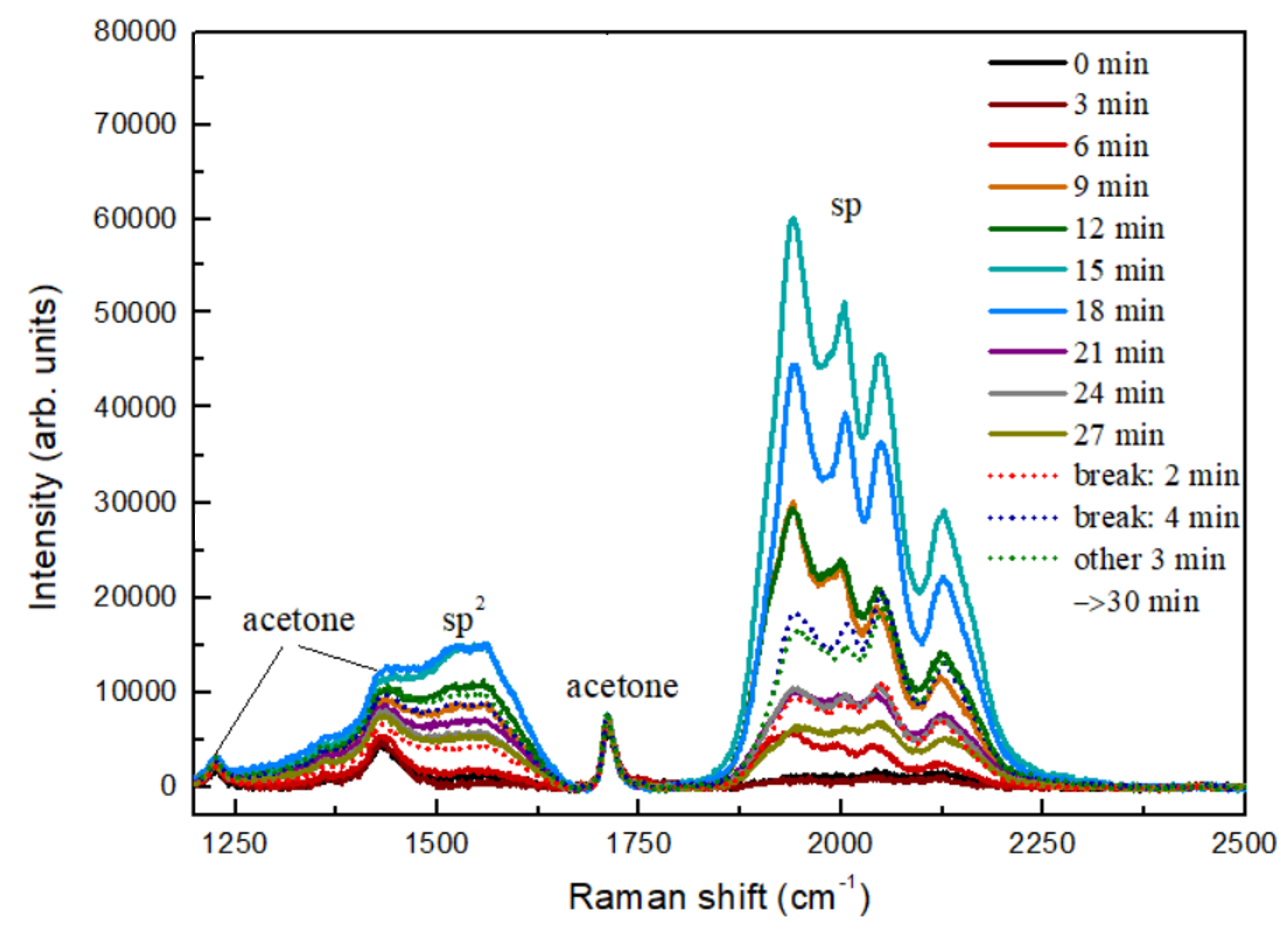}\label{fig:pmma_insitu}}
	\hspace{2mm} 
	\subfloat[][]{\includegraphics[scale=0.53]{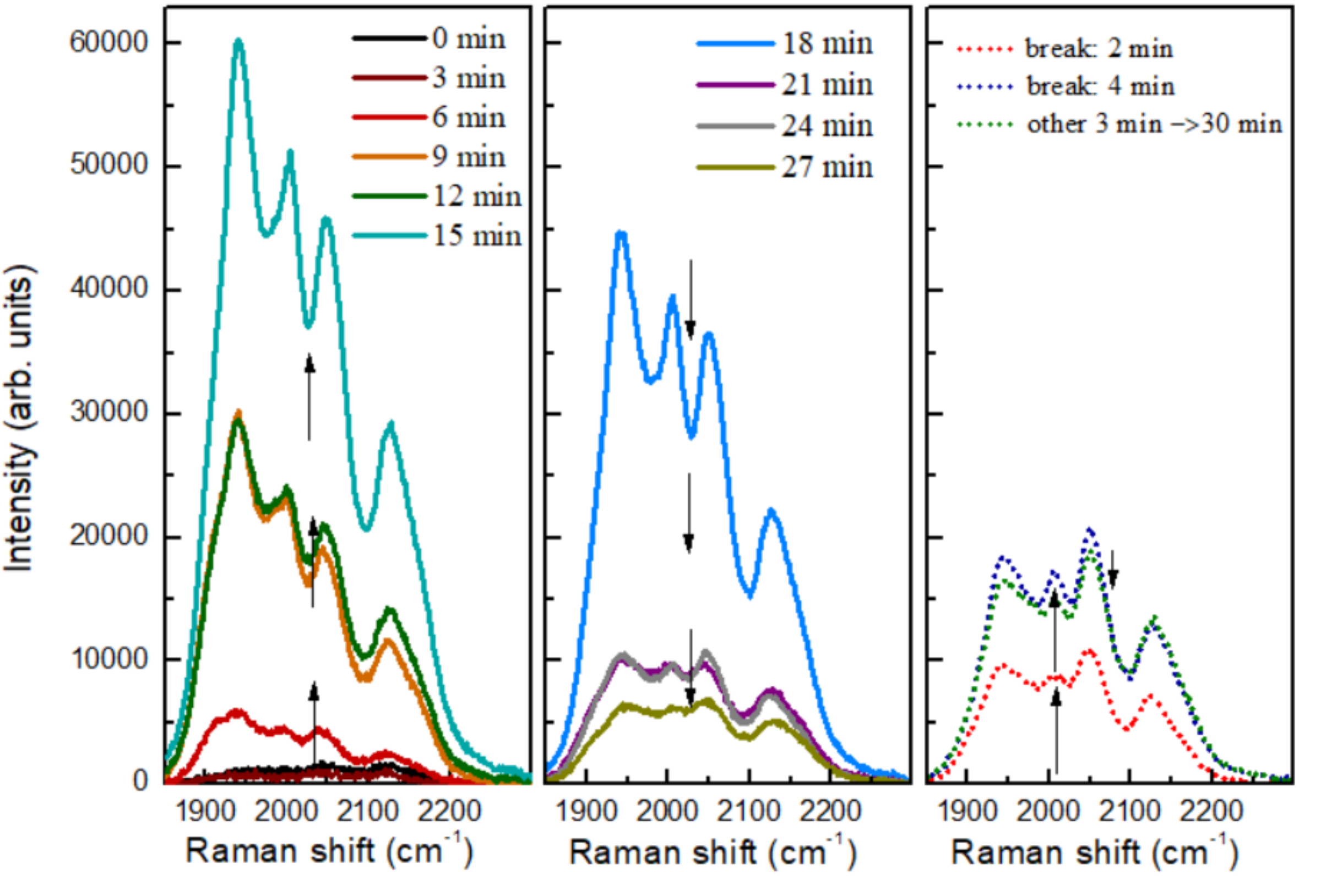}\label{fig:insitu_zoom}}
	\caption[\textit{In situ} SERS measurements (excitation line at 532\,nm) performed between consecutive ablations of graphite in PMMA solution with Ag nanoparticles for a total ablation time of 30 minutes.\protect\subref{fig:pmma_insitu} Spectra with sp$^2$ and sp signals marked. \protect\subref{fig:insitu_zoom} Focus on SERS sp signals in time: increase (left), decrease (centre) and after breaks (right).]{\textit{In situ} SERS measurements (excitation line at 532\,nm) performed between consecutive ablations of graphite in PMMA solution with Ag nanoparticles for a total ablation time of 30 minutes.\protect\subref{fig:pmma_insitu} Spectra with sp$^2$ and sp signals marked.  \protect\subref{fig:insitu_zoom} Focus on SERS sp signals in time: increase (left), decrease (centre) and after breaks (right).} 	
\end{figure}

In the second work, we studied a solution of polyynes in poly(methyl methacrylate) (PMMA) dissoluted in acetone and silver nanoparticles. The formation of sp-carbon chains was characterized by in situ surface-enhanced Raman spectroscopy measurements taken every 3 minutes of ablation, stopping the laser of PLAL, as shown in \cref{fig:pmma_insitu} \cite{tesiPeggiani}. The sp$^2$ and sp carbon signals start to be well defined after 6 minutes of ablation of graphite target. During the ablation, the sp$^2$ band is very broad and does not reach the same intensity of sp signals, which are characterized by four distinct features. To focus on the trend of sp-carbon chains over time, SERS spectra were limited to the sp region as reported in \cref{fig:insitu_zoom}. 
Up to 15 minutes of ablation the sp signal increases and then decreases between 16 and 27 minutes. Additional SERS measurements were performed after 2 and 4 minutes of break, i.e. without ablation, promoting another increase of sp signal. After other 3 minutes of ablation, a further decrease of the sp intensity was observed. 
This peculiar behaviour can be associated to the degradation of polyynes due to the laser irradiation, the saturation of SERS active sites of Ag nanoparticles or to the resizing and reshaping of metal nanoparticles by ns-laser pulses \cite{resize2017}. 
Ag nanoparticles can aggregate and diffuse in the liquid during the break, allowing SERS signals to grow again while, re-irradiating, silver nanoparticles were reshaped promoting once more the decrease of the sp signal.
\\Further development of \textit{in situ} methods similar to the latter will help to better understand the mechanisms of formation of polyynes and also the effect of the different parameters (e.g. laser, target, and solvent) on the synthesis yield of polyynes. In general, ultrafast and high-gain techniques are required to acquire enough information in the shortest timescale possible. Ideally, probes with characteristic times of the order of \textit{fs} to \textit{ns} are required to follow the inner formation of polyynes in the plasma phase. Exploit resonances or inner signal amplification of polyynes can be beneficial to reduce the measurement times. In this sense, the recent advances in synchrotron light sources, ultrafast laser setup, and free electron lasers are encouraging the improvement of \textit{in situ} techniques.

\section{The role of the fluence in the synthesis of polyynes}

Laser fluence is one of the most important parameters concerning the synthesis of nanomaterials through laser ablation. Indeed, the rate of growth of nanomaterials and many of their properties can be directly connected to the choice of the laser fluence employed in ablation experiments. However, the contribution of the fluence to the formation of sp-carbon chains has not been yet investigated in detail. Indeed, the throughput of PLAL experiments with similar conditions seemed to be very different, as reported in the works listed in \cref{solvent_sec} and \cref{target_sec}. Also, the geometry of the ablation process appears to heavily affect the synthesis of sp-carbon chains. Moreover, the detection method employed can determine the ability to observe specific sp-carbon chains.
\\In this framework, we propose a systematic study of the synthesis yield of sp-carbon chains, in particular polyynes, by PLAL in acetonitrile by deeply analyzing the two main parameters that contribute to the fluence, namely the laser energy and the spot size. We have chosen acetonitrile due to the possibility to synthesize a wide range of polyynes both in terms of chain length and terminations \cite{peggianisolventi,waka_2012}. We analyzed the solutions of sp-carbon chains through high-performance liquid chromatography (HPLC), as described in our works \cite{peggianisolventi,marabotti2021situ}. HPLC turned out to be one of the best detection methods to analyze the mixtures of polyynes in the most complete way. Indeed, simple UV-Vis absorption spectroscopy does not allow efficient detection of the different size- and termination-selected polyyne, and the absorption of byproducts makes challenging the estimation of the concentration of polyynes \cite{peggianisolventi,taguchi2017,taguchi2015}. From the experimental point of view, we employed a graphite target of 8 mm in diameter sunk in the glass vial with 2 mL of acetonitrile. The total height of the liquid above the target was approximately 24.7 mm which is an important parameter to estimate the fluence. The laser beam passed through a lens with a focal length of 200 mm and the target-to-lens distance was varied to control the dimension of the spot size. The laser employed for the ablations was a Nd:YAG pulsed laser (Quantel Q-Smart) operating at 10 Hz with a pulse duration of 6 ns. The ablation time was fixed to 15 min for all the experiments. We calculated the fluences following the method illustrated in the Supporting Information of our paper \cite{marabotti2021situ}.
\begin{figure}[!t]
    \centering
    \includegraphics[width=\linewidth, keepaspectratio]{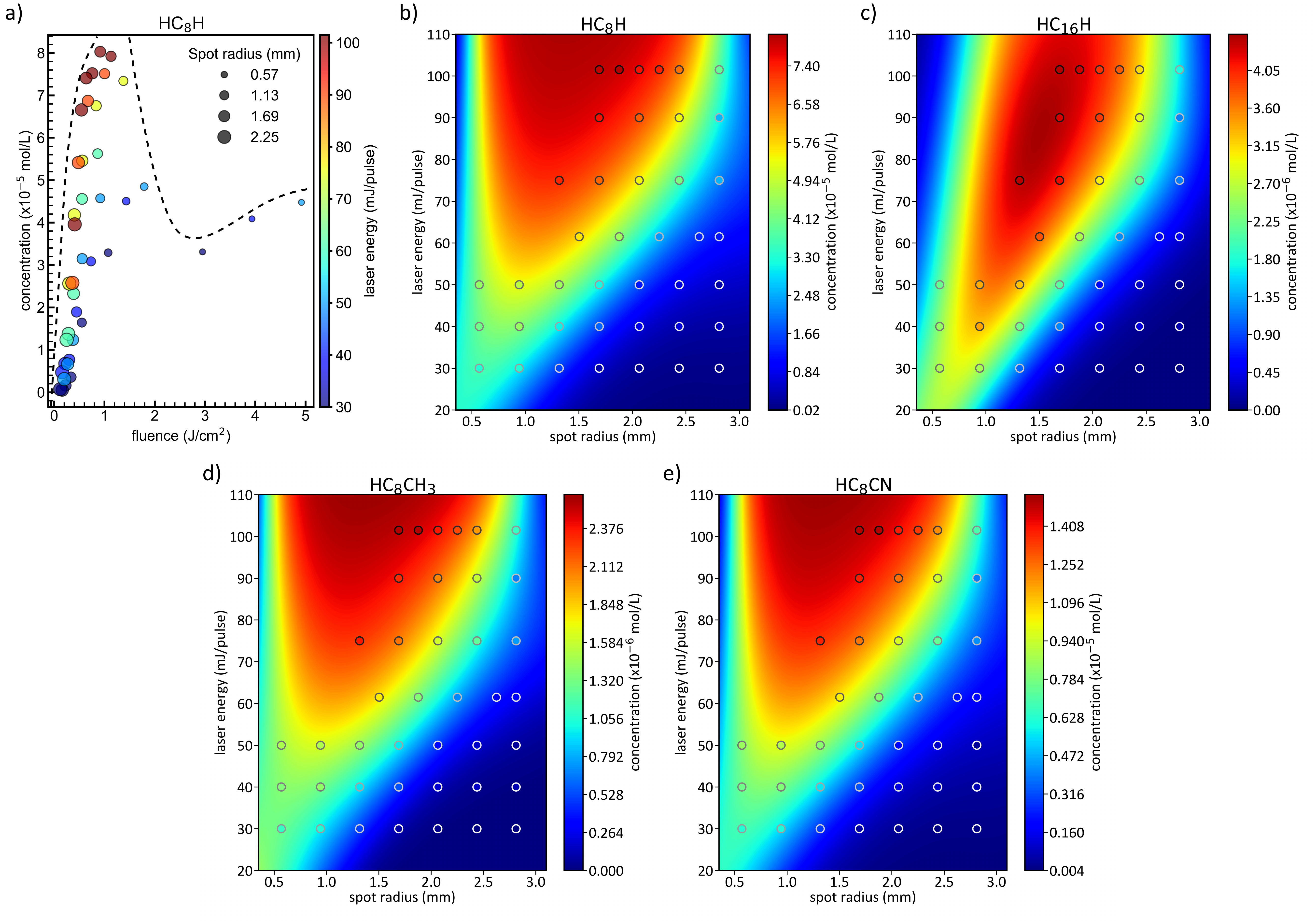}
    \caption{a) Experimental concentration of HC$_8$H extracted from HPLC analysis of polyyne mixtures obtained at different fluences. The color scale follows laser energy, while the dimension of the circles represents the dimension of the spot radius. Experimental concentrations of HC$_8$H in panel b) (colored circles, same as in panel a), HC$_{16}$H in panel c), HC$_8$CH$_3$ in panel d) and HC$_8$CN in panel e) as a function of the spot radius and the laser energy. The different 2D surfaces are obtained from a fit of the experimental points of each corresponding polyyne.}
    \label{fig:fit2D_hydrogen}
\end{figure}
\\We first focused on the synthesis yield of hydrogen-capped polyynes (HC$_n$H) by looking at the fluence as the parameter of merit. \cref{fig:fit2D_hydrogen}a) shows the concentration of HC$_8$H at different fluences that correspond to different combinations of spot radii (size of circles) and laser energies (color of circles). The production of HC$_8$H does not show a clear tendency with the fluence. Indeed, similar values of deposited fluence, related to different pairs of spot radii and laser energies, resulted in a completely different synthesis yield of HC$_8$H. We discovered analogous behaviors for all the detected hydrogen-capped polyynes, \textit{i.e.} from HC$_6$H to HC$_{22}$H. Thus, we cannot rely only on the fluence to predict the synthesis yield of polyynes but we should look singularly at the effect of the spot size and laser energy also to better understand their roles. This observation was already encountered in many studies in which the synthesis yield, shape, and chemistry of nanomaterials are affected in different ways by separately tuning the laser energy or the spot area \cite{guillen2015structure,johny2019sns2,zhang2017colloidal,ganash2019synthesis,reich2017fluence}. Indeed, in ablation experiments, the spot size of the laser beam onto the target plays a simple and intuitive role. Assuming a condition above the ablation threshold of the employed target material, an increase of the spot size increments the amount of material ablated, and thus the synthesis yield is improved as a consequence. Instead, the laser energy can have a more intriguing effect since not always an increase in this parameter coincides with a growth of the synthesis yield.
\\To investigate these parameters in the case of the production of hydrogen-capped polyynes in acetonitrile, we gradually varied the laser energy and the spot size. \cref{fig:fit2D_hydrogen}b) reports the experimental results, \textit{i.e.} the concentration of HC$_8$H, reported by the colored circles, is fitted with an \textit{ad hoc} function to predict the production of HC$_8$H in conditions that have not been tested experimentally, due to the possible presence of dangerous splashes and the flammability of acetonitrile at high fluences (e.g. at small spot radii and high energies). As shown in \cref{fig:fit2D_hydrogen}b), the growth of production yield of HC$_8$H follows a sigmoid-like shape, increasing as the overall fluence increase. However, this behavior seemed to change in the case of sizable fluences featuring small spot radii where we observed a decrease in the synthesis yield both in the experimental and fitted data. The synthesis yield is also varying by increasing the length of the chain as can be observed passing from \cref{fig:fit2D_hydrogen}b), \textit{i.e.} HC$_8$H, to panel c), \textit{i.e.} HC$_{16}$H. Indeed, the concentration of HC$_{16}$H features a maximum at 86.5 mJ/pulse and 1.50 mm of spot radius (fluence of 1.22 J/cm$^2$) and shows a limited region in which the formation of this polyyne is favored.
\\The results reported in \cref{fig:fit2D_hydrogen}b) and c) suggest that high laser energies have beneficial effects if the spot radius is larger than approximately 1.5 mm for quite every chain. It seems that longer hydrogen-capped polyynes are produced with higher efficiency for fluences less than 1 J/cm$^2$, while shorter chains are promoted for increasing fluence. Indeed, the longest polyyne ever produced by PLAL, namely HC$_{30}$H \cite{matsutani2012}, was synthesized by ablation in decalin with a fluence of 0.57 J/cm$^2$. In this sense, we can assume that for higher fluences, other carbon nanostructures can be favored by PLAL synthesis instead of long polyynes \cite{hobley2007formation,amans2017origin}. Moreover, it should exist an optimized concentration ratio between ionized carbon atoms, carbon radicals, and terminating groups (in this case atomic hydrogen) to promote the formation of long polyynes compared to shorter ones or more stable allotropes of carbon. 
\\Despite hydrogen-capped polyynes, we analyzed the synthesis yield of methyl- and cyano-capped polyynes, in \cref{fig:fit2D_hydrogen}d) and e), respectively. In particular, we chose HC$_8$CH$_3$ and HC$_8$CN to compare them with HC$_8$H since they are the most concentrated species of methyl- and cyano-capped polyynes, respectively. As can be noted, the synthesis yields of these two chains follow a trend with the analyzed parameters, \textit{i.e.} laser energy and spot radius, similar to that of the corresponding hydrogen-capped polyyne. We can conclude that the formation mechanism and the influence of the process parameters in the case of short hydrogen-, methyl-, and cyano-capped polyynes are quite similar. However, the concentrations of these species (\textit{i.e.} HC$_8$CH$_3$ and HC$_8$CN) are almost one order of magnitude less than the corresponding hydrogen-capped polyyne (\textit{i.e.} HC$_8$H). These results point out a completely different efficiency of formation, probably connected to a smaller availability of terminating groups as each acetonitrile molecule can provide at most one -CN and one -CH$_3$, while, in principle, it could supply 3 H atoms.

\section{Solvent effect on sp-carbon chains synthesis} \label{solvent_sec}

In this Section, we focus on the effect of the solvent on the synthesis of polyynes by PLAL.
The solvent can indeed have a crucial role in the formation of polyynes by acting as a carbon source (if it is an organic solvent) and by providing terminations \cite{kanitz2019, waka_2012}. 
\\All types of polyynes synthesized by the pulsed laser ablation of a solid graphite pellet in a liquid are reported in \cref{table:solvent}, which is divided into two different macro-groups: experiments performed in aqueous solutions and in organic solvents.

\begingroup
 \footnotesize 
 \centering
 \setlength\LTleft{50pt}%
  \setlength\LTright{50pt}%
    \begin{longtable}{@{\extracolsep{\fill}}|l|l|l|l|l|l@{}|}
    \caption{Polyynes obtained from the ablation of a graphite pellet in different solvents.\label{table:solvent}}\\ 
\hline
    \textbf{Solvent}                      
    & \boldsymbol{$\lambda$}
    & \begin{tabular}[c]{@{}l@{}}\textbf{Freqeuncy},\\ \textbf{Duration}\end{tabular} 
    & \boldsymbol{$\tau$}    
    & \begin{tabular}[c]{@{}l@{}}\textbf{Fluence, Energy Pulse, }\\ \textbf{Spot Size}\end{tabular}
    & \textbf{Polyynes}     
    \\
 \hlineB{2} \multicolumn{6}{|c|} {\textbf{Aqueous Solutions}}\\ \hlineB{2}
 
  water \cite{choi2009}   
                                    & 1064 nm 
                                    & 10 Hz, 5 min 
                                    & 7 ns       
                                    & 5.1 J/cm$^2$, 160 mJ, 2 mm
                                    & C$_n$H$_2$ (n=8, 10) 
      \\\hline
  water \cite{grasso2009}   
                                    & 532 nm  
                                    & 10 Hz, 20 min      
                                    & 5 ns          
                                    & 0.4 J/cm$^2$   
                                    & C$_n$H$_2$ (n=6-10)  
      \\\hline
  water    \cite{compagnini2007,compagnini2008} 
                                    & 532 nm  
                                    & 10 Hz, 20 min       
                                    & 5 ns              
                                    & 0.4 J/cm$^2$ 
                                    & C$_n$H$_2$ (n=6,8)      
      \\ \hline
  water  \cite{peggianisolventi}     
                                    & 532 nm   
                                    & 10 Hz, 15 min    
                                    & 6 ns              
                                    & 2.8 J/cm$^2$
                                    &\begin{tabular}[t]{@{}l@{}}C$_n$H$_2$  (n=6-10),\\ HC$_8$CH$_3$\end{tabular} 
      \\\hline
  water  \cite{forte2013}            
                                    & 532 nm  
                                    & 10 Hz, 20 min    
                                    & 5 ns        
                                    & 0.5 J/cm$^2$        
                                    & C$_n$H$_2$ (n=6, 8)        
    \\ \hline 
 water  \cite{park2013}            
                                    &\begin{tabular}[t]{@{}l@{}}1064 nm;\\ 532 nm; \\ 355 nm; \\ 266 nm  \end{tabular}
                                    & 10 Hz, up to 60 min    
                                    & 5 ns        
                                    & 1.3 J/cm$^2$, 20 mJ-40 mJ       
                                    & \begin{tabular}[t]{@{}l@{}}C$_n$H$_2$  (n=6-10);\\ C$_n$H$_2$  (n=6-10); \\ C$_n$H$_2$  (n=6-10); \\ C$_n$H$_2$  (n=6, 8)  \end{tabular}      
    \\ \hline 
 water at different pH  \cite{shin2011}    
                                       & 1064 nm  
                                       & 10 Hz, 20 min  
                                       & 5 ns        
                                       & 40 mJ, 2 mm  
                                       & C$_n$H$_2$ (n=6-10)   
      \\\hline
  water+D$_2$O \cite{park2012}      
                                       & 1064 nm    
                                       & 10 Hz, 20 min     
                                       & 5 ns          
                                       & 40 mJ, 2 mm  
                                       & C$_n$H$_2$ (n=6-10)   
     \\
  
   \hlineB{2} \multicolumn{6}{|c|}{\textbf{Organic solvents}}\\ \hlineB{2}
  
   acetonitrile \cite{compagnini2007,compagnini2008}  
                                       & 532 nm 
                                       & 10 Hz, 20 min  
                                       & 5 ns        
                                       & 0.4 J/cm$^2$        
                                       & C$_n$H$_2$ (n=6-16)   
      \\\hline
 \begin{tabular}[t]{@{}l@{}} acetonitrile;\\ \\ \\methanol;\\ c-hexane \cite{forte2013} \end{tabular}  
                                       & 532 nm 
                                       & 10 Hz, 20 min   
                                       & 5 ns        
                                       & 0.5 J/cm$^2$     
                                       & \begin{tabular}[t]{@{}l@{}} C$_n$H$_2$ (n=6-14), \\ HC$_n$N (n=7-11),\\ C$_n$N$_2$ (n=6, 8); \\ C$_n$H$_2$ (n=6-12);\\ C$_n$H$_2$ (n=6-12)\end{tabular}                
      \\\hline
 ethanol \cite{khashan_2013} 
                                       & 1064 nm 
                                       & 1 Hz
                                       & 9 ns 
                                       & 20-200 mJ 
                                       & C$_n$H$_2$ (n=6-18)
      \\\hline
 decalin \cite{matsutani2010}        & 532 nm
                                       & 10 Hz, 15-80 min
                                       & 5 ns                    
                                       & 0.31 J/cm$^2$           
                                       & C$_n$H$_2$ (n=6-26)     
     \\\hline
 n-hexane  \cite{matsutani_2008}       
                                       & 532 nm  
                                       & 10 Hz, 5 min   
                                       & 5 ns        
                                       & 0.31 J/cm$^2$       
                                       & C$_n$H$_2$ (n=6-20)    
     \\\hline
 n-hexane \cite{matsutani2009}      
                                       & 532 nm              
                                       & 10 Hz, 180 min  
                                       & 5 ns               
                                       & 0.31 J/cm$^2$           
                                       & C$_n$H$_2$ (n=6-22)    
    \\\hline
 \begin{tabular}[t]{@{}l@{}}methanol;\\  ethanol; \\ 1-propanol; \\ 1-butanol; \\ t-butyl alcohol; \\ n-hexane \cite{matsutani_2011}  \end{tabular} 
                                       & 532 nm              
                                       & 10 Hz, 2.5-5 min   
                                       & 5 ns             
                                       & 22 mJ      
                                       & \begin{tabular}[t]{@{}l@{}}C$_n$H$_2$ (n=6-16);\\ C$_n$H$_2$ (n=6-16);\\ C$_n$H$_2$ (n=6-16);\\ C$_n$H$_2$ (n=6-16);\\ C$_n$H$_2$ (n=6-18);\\ C$_n$H$_2$ (n=6-18);\end{tabular}     
    \\\hline
  \begin{tabular}[t]{@{}l@{}}decalin  \cite{matsutani_2011} \end{tabular} 
                                       & 532 nm  
                                       & 10 Hz, 52.5 min  
                                       & 5 ns          
                                       & 22 mJ    
                                       & \begin{tabular}[t]{@{}l@{}}C$_n$H$_2$ (n=6-28);\end{tabular} 
    \\\hline
 n-hexane \cite{matsutani2011_CC}   
                                      & \begin{tabular}[t]{@{}l@{}}1064 nm;\\ 532 nm;\\ 355 nm\end{tabular} 
                                      & 10 Hz    
                                      & 5 ns   
                                      & 40 mJ  
                                      & \begin{tabular}[t]{@{}l@{}}C$_n$H$_2$ (n=6-22);\\ C$_n$H$_2$ (n=6-22);\\ C$_n$H$_2$ (n=6-20)\end{tabular} 
    \\\hline
 decalin  \cite{matsutani2012}         
                                      & 1064 nm  
                                      & 10 Hz; 180 min  
                                      & 5 ns         
                                      & 0.57 J/cm$^2$  
                                      & C$_n$H$_2$ (n=6-30)   
   \\\hline
  \begin{tabular}[t]{@{}l@{}}n-hexane; \\ c-hexane; \\ n-heptane; \\ n-octane\cite{park2012_alkanes} \end{tabular}  
                                      & 1064 nm    
                                      & 10 Hz; 20 min 
                                      & 5 ns       
                                      & 20 mJ  
                                      & \begin{tabular}[t]{@{}l@{}}C$_n$H$_2$ (n=6-14);\\ C$_n$H$_2$ (n=6-16);\\ C$_n$H$_2$ (n=6-14);\\ C$_n$H$_2$ (n=6-16)\end{tabular}                  
    \\\hline
  \begin{tabular}[t]{@{}l@{}} acetonitrile;\\ \\ \\ \\methanol,\\ ethanol,\\ isopropanol \cite{peggianisolventi}\end{tabular} 
                                     & 532 nm  
                                     & 10 Hz, 15 min   
                                     & 6 ns       
                                     & 2.8 J/cm$^2$ 
                                     & \begin{tabular}[t]{@{}l@{}} C$_n$H$_2$ (n=6-22),\\ HC$_n$CH$_3$ (n=6-18),\\ HC$_n$N (n=7-13);\\ \\C$_n$H$_2$ (n=6-22),\\ HC$_n$CH$_3$ (n=6-18)\end{tabular} 
    \\\hline
liquid N$_2$ \cite{russo2014} 
                                     & 800 nm 
                                     & 1000, 20 min 
                                     & 35 fs 
                                     & 25 J/cm$^2$
                                     & C$_8$N$_2$ 
    \\\hline
 \begin{tabular}[t]{@{}l@{}} acetonitrile;\\ n-hexane;\\ c-hexane \cite{park2012}  \end{tabular}   
                                     & 1064 nm 
                                     & 10 Hz, 20 min  
                                     & 5 ns      
                                     & 40 mJ, 2 mm   
                                     & \begin{tabular}[t]{@{}l@{}} C$_n$H$_2$ (n=6-16);\\ C$_n$H$_2$ (n=6-16);\\ C$_n$H$_2$ (n=6-16)\end{tabular}     
    \\\hline
 TEOS \cite{wu2014}    
                                     & 1064 nm   
                                     & 10 Hz, 5 min   
                                     & 240 $\mu$s    
                                     & 1 J, 0.055 mm 
                                     & \begin{tabular}[t]{@{}l@{}}C$_n$H$_2$ (n=8-16),\\ Si–H doped carbon \\ nanoribbon\end{tabular}   
    \\\hline
  \begin{tabular}[t]{@{}l@{}} isopropanol,\\ methanol,\\ ethanol,\\ decalin, \\tetralin  \cite{tesiBenoliel}\end{tabular}
                                     & 532 nm 
                                     & 10 Hz, 5 min  
                                     & 6 ns      
                                     & 2.8 J/cm$^2$, 50 mJ, 0.75 mm   
                                     &\begin{tabular}[t]{@{}l@{}} \\C$_n$H$_2$ (n=6-20),\\ HC$_n$CH$_3$ (n=8-18)\end{tabular}
    \\
  \hline 
  \end{longtable}
\endgroup

\paragraph{Water and aqueous solutions}\
\\Water is a low-cost and environmentally friendly solvent for the synthesis of polyynes by PLAL which was employed as it is \cite{choi2009,compagnini2007,compagnini2008,grasso2009, peggianisolventi, tesiBenoliel,park2013} or to obtain aqueous solutions \cite{shin2011, park2012, peggianiPVA}.
\\Laser ablation in water has the advantage that avoids the presence of any carbon source apart from that coming from the ablated target (graphite) in the reaction environment \cite{grasso2009}. Moreover, by employing water as a solvent, the contaminant carbon materials including carbon-based byproducts can be greatly reduced compared to the organic solvents \cite{choi2009}. Ablation in water preferentially produces short hydrogen-capped polyynes, maximum of 10 carbon atoms as observed in Refs. \cite{choi2009,compagnini2007,compagnini2008,grasso2009, peggianisolventi, tesiBenoliel}.
\\The formation of longer polyynes in water (i.e. C$_{10}$H$_2$) is helped by employing the lower laser intensity (i.e. 20 mJ) and longer wavelength (i.e. 1064 nm) proved in the work of Park \textit{et al.} \cite{park2013}.
Together with hydrogen-capped polyynes we also observed a type of methyl-capped polyyne, i.e. C$_8$CH$_3$, even if at a much lower concentration than the corresponding hydrogen-capped polyynes ($\sim$ 10$^{-6}$ M) \cite{peggianisolventi}. This phenomenon suggests that terminating the chain by a methyl group is possible thanks to the binding of three hydrogen radicals at one end. 
We observed that water does not contribute to providing carbon atoms for the polyynes formation process and its hydrogen generation rate, which is an important factor in polyynes formation, is the lowest one with respect to the organic solvents \cite{peggianisolventi}. This is justified considering that H–OH bond in water has the highest value (i.e. 4.8 eV) with respect to the molecular bonds in organic liquids, such as C–H (4.3 eV), C–C (3.6 eV), and C–O (3.7 eV) \cite{dean1999, kanitz2019}. The yield of shorter chains can be increased in acidic media for example by adding HCl to water \cite{shin2011}. Indeed, the hydrogen concentration in the plasma plume is of crucial importance to determine which phenomenon prevails between polymerization and hydrogenation \cite{tsuji2003}. It is supposed that water solvates polyynes since the maxima of their absorption bands presented a redshift compared to the theoretical data of UV-Vis spectra regarding the isolated molecules \cite{compagnini2008}. If D$_2$O is added to water, the absorption peaks of polyynes show an apparent blue shift, which increases with the concentration of D$_2$O. This is an index of the change in the electronic structure of the chains induced by D-termination \cite{park2012}.

\paragraph{Organic solvents}\
\\Organic solvents have been extensively employed for PLAL with the intention of obtaining different polyynes in terms of length and terminations: acetonitrile (ACN) \cite{compagnini2008, forte2013, compagnini2007, peggianisolventi, park2012}, liquid N$_2$ \cite{russo2014}, methanol (MeOH) \cite{forte2013,matsutani_2011,peggianisolventi, tesiBenoliel}, ethanol (EtOH) \cite{khashan_2013, matsutani_2011, peggianisolventi, tesiBenoliel}, isopropanol (IPA) \cite{peggianisolventi,tesiBenoliel}, 1-propanol \cite{matsutani_2011}, 1-butanol \cite{matsutani_2011}, t-butyl alcohol \cite{matsutani_2011}, c-hexane \cite{forte2013,park2012_alkanes,park2012}, n-hexane \cite{matsutani_2008, matsutani2009,matsutani_2011,matsutani2011_CC,park2012_alkanes, park2012}, n-heptane \cite{park2012_alkanes}, n-octane \cite{park2012_alkanes}, TEOS \cite{wu2014}, decalin \cite{matsutani2010, matsutani_2011, matsutani2012, tesiBenoliel} and tetralin \cite{tesiBenoliel}.
\\Organic solvents are decomposed by the absorption of laser energy and undergo different rates of photolysis and pyrolysis. Each solvent employed in literature for pulsed laser ablation in liquid will be discussed in the following.
\\Acetonitrile highly degrades during laser ablation and has a higher tendency
towards carbonization compared to alcohols \cite{cataldotetra2004}. Indeed, the C/H ratio
of ACN is equal to 0.66, while it is 0.33 for MeOH, 0.4 for EtOH, and 0.5 for IPA. The promoted coking reaction of acetonitrile helps in the formation of polyynes thanks to the presence of carbon species in the plasma region. This allows reaching a concentration of 1.42\,x\,10$^{-4}$ M for the case of polyyne 8 carbon atoms long. However, also the concentration of carbon-based byproducts is higher and this leads to higher UV-Vis background and lower index of purity (i.e. ratio between polyynes and byproducts absorption defined by Taguchi \textit{et al.} \cite{taguchi2015}) as we described in Ref.\cite{peggianisolventi}. In this paper, we also discussed that acetonitrile is a good environment for polyynes because has the slightest value of oxygen dissolved and the lowest polarity. In general, solvents with low polarity and a low quantity of oxygen dissolved help in preserving sp-carbon chains, which are nonpolar molecules and are likely to oxidize in presence of oxygen.
Pulsed laser ablation in acetonitrile of a graphite pellet led to the formation of hydrogen-capped polyynes with a maximum length of 14 \cite{forte2013}, 16 \cite{compagnini2008, compagnini2007, park2012}, and 22 \cite{peggianisolventi} carbon atoms. In addition to hydrogen-capped polyynes, it was possible to observe after the ablation in acetonitrile methyl-capped polyynes up to 18 carbon atoms \cite{peggianisolventi}, cyano-capped polyynes up to 11 carbon atoms \cite{forte2013}, and in our work up to 13 \cite{peggianisolventi} and dicyano-capped polyynes up to 8 carbon atoms \cite{forte2013}. This is an evidence of the degradation of the liquid due to the high plasma temperature in molecular radicals (i.e. methyl and cyano groups), which can terminate the sp-carbon growing chain \cite{peggianisolventi}. We observed methyl-groups also in the case of PLAL in water, which is not characterized by such functional groups. This means that also single atomic radicals can bind together forming molecular structures, as the methyl group, able to terminate the sp-carbon chains  \cite{peggianisolventi}.
Ref. \cite{russo2014} showed that the fs-ablation of graphite in liquid nitrogen induces the formation of dicyano-capped polyynes long 8 carbon atoms. The species was found only in specific conditions of fluence, i.e. in the range between 20 and 30 J/cm$^{2}$, where N$_2$ can be ionized and attach to polyynes, thus forming cyanopolyynes. 
\\ When we used methanol as a solvent for graphite ablation we synthesized hydrogen-capped polyynes with a maximum length of 20 carbon atoms \cite{tesiBenoliel} while in other literature works lengths up to 16 and 12 were detected in Refs. \cite{matsutani_2011,forte2013}, respectively. In the paper of Forte \textit{et al.}, it was observed that polar solvent, like methanol, increases the negative charge along the chain inducing a positive charge onto hydrogen atoms \cite{forte2013}. As the chain increases in length, the net charge in the central region tends to  become neutral. In the focus region, the high energy density is able to promote the formation of carbon chain anions that are immediately neutralized by the acidity of polar solvent \cite{forte2013}. Methanol resulted as the solvent for PLAL by which the polyynes solution has fewer byproducts compared to acetonitrile, ethanol, and isopropanol and is able to allow hydrogen-capped polyynes up to 22 carbons, as reported in our work \cite{peggianisolventi}. Moreover, we also analyzed methyl-capped polyynes up to 18 carbon atoms \cite{peggianisolventi,tesiBenoliel}.
\\Employing ethanol as a solvent for PLAL, hydrogen-capped polyynes up to 16 carbon atoms were found in Ref. \cite{matsutani_2011} and up to 18 in Ref. \cite{khashan_2013}. With the same solvent, we were able to reach a length of 22 carbon atoms for hydrogen-capped polyynes \cite{peggianisolventi} and methyl-capped polyynes up to 18 carbon atoms \cite{peggianisolventi, tesiBenoliel}. We also showed that ethanol ensures slightly greater stability for polyynes than isopropanol. This effect is presumably linked to the lower quantity of oxygen dissolved in ethanol \cite{peggianisolventi}.
\\Other alcohols such as 1-propanol, 1-butanol, and t-butyl alcohol were studied as solvents for PLAL by Matsutani \textit{et al.} \cite{matsutani_2011}, from which hydrogen-capped polyynes up to 16 carbons for the first two solvents and 18 carbons in the case of t-butyl alcohol were synthesized. Matsutani \textit{et al.} \cite{matsutani_2011} did not observe the effect of the solvent polarity on polyynes formation when a graphite pellet is ablated. With a different set of solvents, we observed an increasing yield of polyynes obtained by ablating graphite pellet as the solvent polarity decreases \cite{peggianisolventi}. Other solvent properties which play a crucial role in the polyynes yield can be also: oxygen dissolved, carbonization tendency, C/H ratio, and thermal conductivity. 
We widely explored the effect of the C/H ratio of the solvent on polyynes yield \cite{tesiBenoliel}. We observed an overall decrease in short polyynes from low to high C/H solvent (i.e. 0 for water, 0.33 for MeOH, 0.4 for EtOH, and 0.49 for IPA), as well as an increase in the concentration of longer ones.
\\Cyclohexane (i.e. c-hexane) was used to obtain hydrogen-capped polyynes up to 12, 14, and 16 carbon atoms in Refs. \cite{forte2013,park2012_alkanes,park2012}, respectively.
In Ref.\cite{forte2013}, c-hexane shows a saturable absorption mechanism due to the high energy density which is able to promote the formation of carbon chain anions in the focus region. They are not visible in the UV-Vis spectrum because their lifetime is limited to the laser pulse. 
\\N-hexane was widely employed for the production of polyynes by PLAL, reaching hydrogen-capped polyynes as long as 14 carbons \cite{park2012_alkanes}, 16 \cite{park2012}, 18 \cite{matsutani_2011}, 20 \cite{matsutani_2008}, 22 \cite{matsutani2009, matsutani2011_CC}. The higher concentration of polyynes synthesized in n-hexane compared to water can be linked to the ease of fragmentation of carbon-carbon bonds of the n-hexane. The fragmented species interact with the graphitic plasma leading to longer polyynes \cite{park2012}.
\\In the work of Park \textit{et al.}, properties of different solvents (i.e. c-hexane, n-hexane, n-heptane n-octane) as bond dissociation energy, thermal conductivity, and total mass of hydrogen atoms per volume of solvent are not affecting the production of polyynes \cite{park2012_alkanes}. Conversely, the polyynes yield is strongly affected by the ratio of the hydrogen and carbon atoms in the solvent molecules. The maximum length of the hydrogen-capped polyynes presented in this work is 16 in the case of c-hexane and 14 for the other alkanes.
\\Pulsed laser ablation in liquid was also performed using tetraethyl orthosilicate (TEOS) as a solvent and graphite pellet as a target, finding hydrogen-capped polyynes up to 16 carbon atoms \cite{wu2014}. Graphite pellet and TEOS provide proton and carbon species to the laser plume for polyynes production, which takes place following four steps: 
\begin{enumerate}
\item Decomposition of TEOS
\item Generation of C$_2$ radicals from both TEOS and graphite target
\item mC$_2$ (from target, from reaction 1.) and nC$_2$ (from solvent, from reaction 2.)= (C$_2$)$_{m+n}$ 
\item (C$_2$)$_{m+n}$ + 2H {\normalfont$\rightarrow$  H(C$_2$)$_{m+n}$H}
\end{enumerate}
Together with polyynes, PLAL of graphite in TEOS generated also carbon nanoplates/ribbons composed of graphene-based lamellae with Si–H dopant \cite{wu2014}.
\\The longest hydrogen-capped polyynes ever produced so far by PLAL was 30 atoms of carbon long \cite{matsutani2012}. This was possible after finding the optimized PLAL parameters for the production (i.e. fluence, wavelength, ablation time, glass bottle diameter) and the optimized HPLC parameters for the separation. A fundamental parameter to obtain the longest polyynes by PLAL was the choice of decalin as a solvent. Indeed, the ablation in decalin allows the formation of long polyynes also in other literature works: 20 carbon atoms in Ref. \cite{tesiBenoliel}, 26 in Ref. \cite{matsutani2010}, and 28 in Ref. \cite{matsutani_2011}. We detected methyl-capped polyynes up to 18 atoms of carbon \cite{tesiBenoliel}. Decalin has a viscosity coefficient (i.e. 3.042 mPa s) higher than the other solvents \cite{matsutani2010}, so the density of C$_2$ radicals in the plasma region may be higher due to the suppression of radical diffusion. This effect may hinder the termination reaction on both ends of polyynes promoting the growth of the chain \cite{matsutani2010}. In addition, decalin has a relatively low number of hydrogen atoms compared to the other studied solvents (i.e. its C/H ratio is equal to 0.56), so the termination reaction on both ends of the sp-carbon chains hardly occurs leading to the synthesis of long-chain polyynes. 
\\When we employed tetralin as a solvent for PLAL, which has an higher C/H ratio (i.e. 0.83) compared to that of decalin, we observed an overall decrease of short polyynes, as well as an increase in the concentration of longer ones \cite{tesiBenoliel}.
Decalin and tetralin could ideally contribute to the chain growth thanks to 10 consecutive C-C bonds, assuming a complete opening of the hexagonal rings. The concentration of long polyynes has been further related to an increase in viscosity and a decrease in polarity of the solvent.
\\Knowledge about the solvent effect on sp-carbon chains synthesis by PLAL is important since this gives valuable information about the tunability of polyynes in terms of length and termination. Although solvent decomposition during laser ablation has been reported in the mentioned literature works, the detailed impact of the liquid media on polyynes synthesis by PLAL needs to be further investigated.

\section{Target effect on sp-carbon chains synthesis} \label{target_sec}
In this Section, we focus on the effect of target on the synthesis of polyynes by PLAL. The target, indeed, heavily affects all the ablation process since it determines the fundamental interaction with the laser pulses and all the cascade of optical and thermal phenomena that follows it \cite{kanitz2019,amendola2013}.
\\A large variety of target has been adopted for the synthesis of polyynes by PLAL which are listed in \cref{table:target}, apart from graphite pellet which was discussed in the previous Section. This Section is divided based on the kind of target employed: ablations with solid targets, powders and without target, \textit{i.e.} by simply irradiating the pure solvents with femtosecond laser pulses.


\begingroup
  \footnotesize 
  \centering
    \begin{longtable}{@{\extracolsep{\fill}}|p{3 cm}|l@{ }|l@{ }|l@{ }|l@{ }|l@{ }|l@{ }|@{}} 
          \caption{Polyynes by ablating different targets in liquid.\label{table:target}}\\ 
\hline
    \textbf{Target}    
    & \textbf{Solvent}
    & \boldsymbol{$\lambda$} 
    & \begin{tabular}[c]{@{}l@{}}\textbf{Frequency},\\ \textbf{Duration}\end{tabular}
    & \boldsymbol{$\tau$}                                        
    & \begin{tabular}[c]{@{}l@{}}\textbf{Fluence, Energy Pulse},\\ \textbf{ Spot Size}\end{tabular}
    & \textbf{Polyynes}
    \\ 
\hlineB{2} \multicolumn{7}{|c|} {\textbf{Solid Targets}}\\ \hlineB{2}

shungite \cite{kuche2016}   
                            & water   
                            & 1064 nm     
                            & 50 Hz     
                            & 2 ms    
                            & 1-3 J     
                            & SERS Sp-band       
    \\\hline
PTCDA pellet  \cite{matsutani_2008} 
                            & n-hexane  
                            & 532 nm  
                            & 10 Hz, 5 min  
                            & 5 ns 
                            & 0.31 J/cm$^2$
                            & C$_n$H$_2$ (n=6-18) 
    \\ \hline
PTCDA pellet  \cite{matsutani_2011} 
                            & \begin{tabular}[t]{@{}l@{}}methanol; \\ ethanol; \\ 1-propanol; \\ 1-butanol; \\ t-butyl alcohol; \\ n-hexane\end{tabular}            
                            & 532 nm                       
                            & 10 Hz; 30-45 min       
                            & 5 ns         
                            & 22 mJ       
                            & \begin{tabular}[t]{@{}l@{}}C$_n$H$_2$ (n=6-10);\\ C$_n$H$_2$ (n=6-10);\\ C$_n$H$_2$ (n=6-10);\\ C$_n$H$_2$ (n=6-16);\\ C$_n$H$_2$ (n=6-18);\\C$_n$H$_2$ (n=6-18)\end{tabular}  
    \\ \hline
fullerene pellet  \cite{matsutani2011_CC} 
                            & n-hexane          
                            & \begin{tabular}[t]{@{}l@{}}1064 nm;\\ 532 nm;\\ 355 nm\end{tabular}    
                            & 10 Hz      
                            & 5 ns     
                            & 40 mJ    
                            & \begin{tabular}[t]{@{}l@{}}C$_n$H$_2$ (n=6-22);\\ C$_n$H$_2$ (n=6-22);\\ C$_n$H$_2$ (n=6-20)\end{tabular}   
    \\\hline
fullerene pellet  \cite{matsutani2011_CC} 
                            & methanol        
                            & \begin{tabular}[t]{@{}l@{}}1064 nm;\\ 532 nm;\\ 355 nm\end{tabular}   
                            & 10 Hz
                            & 5 ns        
                            & 40 mJ        
                            & C$_n$H$_2$ (n=6-20)       
    \\\hline 
naphthalene pellet  \cite{tesiBenoliel}
                            &\begin{tabular}[t]{@{}l@{}} water;\\ isopropanol,\\ methanol,\\ ethanol \end{tabular}
                            & 532 nm 
                            & 10 Hz, 5 min  
                            & 6 ns      
                            & 2.8 J/cm$^2$, 50 mJ, 0.75 mm   
                            &\begin{tabular}[t]{@{}l@{}} C$_n$H$_2$ (n=6-10);\\ C$_n$H$_2$ (n=6-18),\\ HC$_n$CH$_3$ (n=8-16)\end{tabular}
    \\
    
\hlineB{2} \multicolumn{7}{|c|} {\textbf{Powder Targets}}\\ \hlineB{2}


graphite particles  \cite{aru2015} 
                            & \begin{tabular}[t]{@{}l@{}}water; \\ n-hexane\end{tabular}         
                            & 800 nm                   
                            & 1000 Hz   
                            & 160 fs     
                            & (0.5, 1) mJ  
                            & \begin{tabular}[t]{@{}l@{}}C$_n$H$_2$ (n=4-12); \\ C$_n$H$_2$ (n=4-20)\end{tabular}   
    \\ \hline
graphite particles   \cite{choi2009}    
                            & water      
                            & 1064 nm       
                            & 10 Hz, 5 min   
                            & 7 ns       
                            & 5.1 J/cm$^2$, 160 mJ, 2 mm        
                            & -             
    \\ \hline
\begin{tabular}[t]{@{}l@{}}graphite powder;\\ PTCDA powder\cite{matsutani_2008}\end{tabular}   
                            & n-hexane          
                            & 532 nm        
                            & 10 Hz, 5 min 
                            & 5 ns      
                            & 0.31       
                            & \begin{tabular}[t]{@{}l@{}}C$_n$H$_2$ (n=6-20);\\ C$_n$H$_2$ (n=6-18)\end{tabular}  
    \\ \hline
graphite particles \cite{nishide2006, waka2007CPL} 
                            & n-hexane          
                            & 532 nm          
                            & 10 Hz, 60 min  
                            & 5 ns       
                            & 100 mJ, 100 mm       
                            & C$_n$H$_2$ (n=8-12)     
    \\ \hline
diamond nanoparticles  \cite{tabata2004}   
                            & ethanol                  
                            & 532 nm            
                            & 20 Hz, 0-120 min  
                            & 7 ns        
                            & 1.3 J/cm$^2$, 2 mm      
                            & C$_n$H$_2$ (n=8-16) 
    \\ \hline
 diamond nanoparticles  \cite{tabata2006_long}   
                            & ethanol                  
                            & 532 nm            
                            & 20 Hz, 0-160 min  
                            & 7 ns        
                            & 1.3 J/cm$^2$      
                            & C$_n$H$_2$ (n=8-20) 
    \\ \hline
\begin{tabular}[t]{@{}l@{}}diamond nanoparticles;\\ carbon onions;\\ graphite powder \cite{tabata2005}\end{tabular} 
                            & ethanol       
                            & 532 nm        
                            & 20 Hz, 5-10 min 
                            & 7 ns        
                            & 1.3 J/cm$^2$, 2 mm     
                            & \begin{tabular}[t]{@{}l@{}}C$_n$H$_2$ (n=8-16);\\ C$_n$H$_2$ (n=8-14);\\ C$_n$H$_2$ (n=8-14)\end{tabular}   
    \\ \hline
graphite powder \cite{tabata2006}   
                            & ethanol      
                            & 532 nm         
                            & 10 Hz, 180 min  
                            & 5 ns          
                            & 0.21 mJ, 9 mm        
                            & C$_n$H$_2$ (n=8-16)    
    \\ \hline
graphite powder   \cite{tsuji2002} 
                            & benzene          
                            & 355 nm        
                            & 10 Hz       
                            & 5-9 ns     
                            & 200 J/cm$^2$, 40 mJ      
                            & C$_n$H$_2$ (n=10-16)        
    \\ \hline
graphite powder \cite{tsuji2002}  
                            & toluene       
                            & \begin{tabular}[t]{@{}l@{}}1064 nm;\\ 532 nm;\\ 355 nm\end{tabular}   
                            & 10 Hz      
                            & \begin{tabular}[t]{@{}l@{}}1 $\mu$s;\\ 5-9 ns;\\ 5-9 ns\end{tabular}
                            & 200 J/cm$^2$, 40 mJ           
                            & \begin{tabular}[t]{@{}l@{}}C$_n$H$_2$ (n=10-14);\\ C$_n$H$_2$ (n=10-16);\\ C$_n$H$_2$ (n=10-16)\end{tabular} 
    \\ \hline
graphite powder  \cite{tsuji2002}     
                            & n-hexane     
                            & 355 nm        
                            & 10 Hz      
                            & 5-9 ns        
                            & 200 J/cm$^2$, 40 mJ   
                            & C$_n$H$_2$ (n=8-14)     
    \\ \hline
fullerene particles \cite{tsuji2003}   
                            & n-hexane     
                            & \begin{tabular}[t]{@{}l@{}}1064 nm;\\532 nm;\\ 355 nm;\\ 266 nm\end{tabular} 
                            & 10 Hz, 40-120 min   
                            & 5-9 ns       
                            & 200 J/cm$^2$, 20-80 mJ      
                            & \begin{tabular}[t]{@{}l@{}}C$_8$H$_2$;\\ C$_n$H$_2$ (n=8,10);\\ C$_n$H$_2$ (n=8-12);\\ C$_8$H$_2$\end{tabular}   
    \\ \hline
fullerene particles \cite{tsuji2003}  
                            & methanol       
                            & \begin{tabular}[t]{@{}l@{}}1064 nm;\\ 532 nm;\\ 355 nm;\\ 266 nm\end{tabular} 
                            & 10 Hz, 40-60 min 
                            & 5-9 ns       
                            & 200 J/cm$^2$, 20-80 mJ      
                            & \begin{tabular}[t]{@{}l@{}}-;\\ -;\\ C$_n$H$_2$ (n=8,10);\\ C$_8$H$_2$\end{tabular}     
    \\ \hline
graphite particles  \cite{wada2012} 
                            & \begin{tabular}[t]{@{}l@{}}n-hexane + iodine;\\ CCl$_4$ + iodine\end{tabular}      
                            & 1064 nm                     
                            & 10 Hz    
                            & 5 ns        
                            & 800 mJ       
                            & \begin{tabular}[t]{@{}l@{}}C$_n$H$_2$I$_6$ (n=10-18);\\ C$_n$H$_2$I$_6$ (n=10-14)\end{tabular} 
    \\ \hline
graphite powder \cite{waka_2012}  
                            & acetonitrile       
                            & 532 nm           
                            & 10 Hz, 300 min 
                            & 5 ns    
                            & 0.4 J           
                            & \begin{tabular}[t]{@{}l@{}}C$_n$H$_2$ (n=8-14),\\ HC$_n$N (n=7-13)\end{tabular} 
    \\ 
   \hlineB{2} \multicolumn{7}{|c|}{\textbf{Direct Ablation of the Solvent}}\\ \hlineB{2}
 

no target     \cite{hu2008}   
                            & acetone       
                            & 800 nm       
                            & 1000 Hz    
                            & 90 fs     
                            & 300 $\mu$J, 10 $\mu$n      
                            & C$_6$H$_2$       
    \\ \hline
no target  \cite{rama2015, rama2017}     
                            & toluene   
                            & 800 nm           
                            & 1000 Hz     
                            & 35 fs    
                            & 300 $\mu$J        
                            & \begin{tabular}[t]{@{}l@{}}C$_n$H$_2$ (n=12-18);\\ HC$_n$CH$_3$ (n=12,14)\end{tabular}
    \\ \hline
no target  \cite{sato2010}           
                            & \begin{tabular}[t]{@{}l@{}}n-hexane;\\ decane\end{tabular}     
                            & 800 nm               
                            & 1000 Hz, 30-300 min 
                            & 100 fs    
                            & 0.9 mJ     
                            & \begin{tabular}[t]{@{}l@{}}C$_n$H$_2$ (n=8-12);\\ C$_n$H$_2$ (n=8-10);\end{tabular}  
    \\ \hline
no target \cite{weso2011}      
                            & benzene       
                            & 800 nm         
                            & 1000 Hz, 90 min 
                            & 120 fs       
                            & 300 $\mu$J       
                            & C$_n$H$_2$ (n=8,14)   
    \\ \hline
no target  \cite{zaidi2010,zaidi2019femtosecond}            
                            & n-octane         
                            & 800 nm          
                            & 100 Hz, 120 min 
                            & 100 fs        
                            & 300 $\mu$J     
                            & C$_n$H$_2$ (n=8-14)   
    \\
    \hline 
  \end{longtable}
\endgroup

\paragraph{Solid targets}\
\\Besides graphite pellet, other solid targets have been used to produce polyynes by PLAL as shungite \cite{kuche2016}, PTCDA pellet \cite{matsutani_2008,matsutani_2011}, fullerene pellet \cite{matsutani2011_CC}, and naphthalene pellet \cite{tesiBenoliel}. 
\\In the work of Kuche \textit{et al.} \cite{kuche2016}, shungite has been used as the solid target for the ablation in colloidal solution with carbon, silver, and/or gold nanoparticles. The authors observed the presence of the characteristic vibrational bands of sp-carbon chains in surface-enhanced Raman scattering spectra. These bands change shape and intensity by changing the concentration of metal nanoparticles dissolved in the solution before the ablation, thus pointing out that the structure of polyynes and cumulenes synthesized in the solution is influenced by the concentration of these nanostructures \cite{kuche2016}.
\\PTCDA pellets were employed as the target for ablation in different solvents \cite{matsutani_2008,matsutani_2011}. Hydrogen-capped polyynes were efficiently synthesized up to 10 carbon atoms in methanol, ethanol, and 1-propanol \cite{matsutani_2011}, 16 in 1-butanol \cite{matsutani_2011}, 18 in t-butyl alcohol, and n-hexane \cite{matsutani_2008,matsutani_2011}. The production of polyynes, both in terms of maximum chain length and concentration, with PTCDA pellets is lower compared to graphite pellets when ablating in n-hexane \cite{matsutani_2008}. Indeed, the presence of carboxylic dianhydride at the side of PTCDA molecules reduces the total amount of C$_2$ radicals that can be formed by the action of the laser pulses \cite{matsutani_2008}. Moreover, these side groups easily transfer the laser energy received to the surrounding molecules of polar solvents \textit{via} electrostatic interaction and hydrogen bonding, hindering once more the formation of C$_2$ radicals \cite{matsutani_2011}. For this reason, the production of polyynes with PTCDA pellet is maximized when nonpolar solvent are employed, as n-hexane.
\\Matsutani \textit{et al.} ablated fullerene pellets in methanol and n-hexane with different harmonics of the Nd:YAG laser, namely at 1064, 532, and 355 nm. They observed an efficient production of polyynes up to 20 carbon atoms in methanol and 22 in n-hexane \cite{matsutani2011_CC}. However, the overall polyyne concentration and the production of long-polyynes decrease passing from 1064 to 355 nm laser wavelength. This behavior is justified by considering that fullerene particles, that can be released during the ablation, start absorbing below 600 nm, reducing the amount of energy that reaches the fullerene pellet by employing 532 and 355 nm lasers \cite{matsutani2011_CC}.
\\We were the first ones in employing naphthalene pellet as an ablation target for PLAL in water and alcohols (methanol, ethanol, and isopropanol) \cite{tesiBenoliel}. The ablation in water proved that naphthalene pellet can be employed for the production of polyynes since water does not act as a carbon source. Moreover, the amount of polyynes produced with the naphthalene target in water is larger compared to ablation with a graphite target in the same conditions. We found the same results also in alcohols and, in particular, the production of long polyyne featured a sizeable increase compared to the use of a graphite pellet. Indeed, it was suggested that the aromatic rings of naphthalene can be opened due to the action of the laser pulses and furnish long carbon radicals, from C$_2$ to maximum C$_{10}$, that promote the formation of longer polyynes \cite{tesiBenoliel}.

\paragraph{Powder targets}\
\\The powder targets employed to synthesize polyynes by PLAL were of a different nature: graphite particles \cite{aru2015, choi2009, nishide2006, waka2007CPL, wada2012},  diamond nanoparticles \cite{tabata2004, tabata2006_long, tabata2005}, carbon onions \cite{tabata2005}, graphite powder \cite{matsutani_2008, tabata2005, tabata2006, tsuji2002, waka_2012}, fullerene particles \cite{tsuji2003} and PTCDA powder \cite{matsutani_2008}.
\\Graphite particles, under laser irradiation, undergo the break of carbon-carbon links obtaining single ionized atoms and the plasma formation. Carbon atoms start to assemble into chains and terminate with H atoms \cite{aru2015}.
The work of Aru \textit{et al.} showed the formation of hydrogen-capped polyynes of 12 atoms long in the case of ablation of graphite particles in water and 20 carbon atoms if n-hexane is employed \cite{aru2015}. The work of Choi \textit{et al.}, employing  water as a solvent and graphite particles and graphite pellet as targets, observed that polyynes were only formed in the case of the graphite pellet \cite{choi2009}. Hydrogen-capped polyynes up to 12 carbons were obtained in n-hexane ablating graphite particles in Refs. \cite{nishide2006,waka2007CPL}.
\\Diamond nanoparticles 5 nm in diameter can successfully be used as a target for the production of polyynes by laser ablation in ethanol \cite{tabata2004}. Compared to when graphite particles are ablated, the amount of long polyynes (up to 16 carbon atoms) is larger with diamond nanoparticles. 
Those results suggest a two-fold process of polyynes formation: the direct evaporation of carbon atoms from diamond nanoparticles and the subsequent evaporation of carbon atoms from graphitic nanoparticles transformed from initial diamond nanoparticles by laser ablation \cite{tabata2004}.
It was showed in Ref. \cite{tabata2005} that the chain length distribution of the polyynes produced from diamond nanoparticles depends on the ablation time. Shorter time induces larger fractions of long polyynes, which corresponds to polyynes produced by the direct evaporation of carbon atoms from initial diamond nanoparticles. The relative fraction of long polyynes decreases with the irradiation time.
\\The chain length distributions of the polyynes produced by laser ablation in ethanol with carbon onions as targets is very similar to the case of the ablation of graphite powder. This result shows that the chain length distribution of the sp-carbon chains using graphite sources is intrinsic and independent from the particle size \cite{tabata2005}. Conversely, in the case of diamond nanoparticles, the chain length distribution of polyynes depends on the size of the particles \cite{tabata2006_long}. The amounts of short polyynes with 6 and 8 carbons are larger when they are produced from larger diamond nanoparticles. On the other hand, the amounts of long polyynes such as C$_{14}$H$_2$ and C$_{16}$H$_2$  are larger when they are produced from smaller ones. Since short polyynes constitute the majority of polyyne in the mixture, the total production yield of sp-carbon chains is therefore larger in the case of diamond nanoparticles with larger diameter \cite{tabata2006_long}.
\\Tsuji \textit{et al.} employed a target of graphite powder dispersed in different solvents (i.e. benzene, toluene, n-hexane) for laser ablation in liquid, which was the first time that polyynes were synthesied by PLAL \cite{tsuji2002}. Different wavelengths were exploited (i.e. 355, 532, 1064 nm) for polyynes synthesis. Hydrogen-capped polyynes of different length were produced in this way reaching the maximum length of 16 carbon atoms. It was claimed that the chain length is independent from the energy density of the radiation and there is an optimum yield of polyynes at specific concentration of graphite powder in the solvent. When the graphite dispersed in the liquid is too high, scattering of the laser can happen leading to a decrease in the efficiency of the process. Moreover, the formation of polyynes seemed to decrease with increasing the laser wavelength, even if the synthesis of polyynes was achieved in all the cases \cite{tsuji2002}.
Conversely, a work of Matsutani \textit{et al.}  declared that higher yield of polyynes were observed at higher laser wavelength \cite{matsutani2011_CC}. This effect can be explained considering the different type of target employed. If a solid pellet is located at the bottom of the cell, instead of graphite powder dispersed in the solvent, the laser beam crosses the whole liquid volume and can be scattered or absorbed from the by-products produced during the ablation. This phenomenon is enhanced if the wavelength is short, i.e. 355 nm, which is closer to the absorption of impurities and to the absorption of long polyynes, that may be broken from the laser irradiation.
In the work of Matsutani \textit{et al.}  graphite and PTCDA powders were dispersed in n-hexane and the results were compared to the case of graphite and PTCDA pellet, respectively \cite{matsutani_2008}. The amount of short polyynes (i.e. C$_8$H$_2$) was four times larger in the case of powder, while the longer chains up to 20 carbons are relatively less produced in the case of the target. In general, the total amount of polyynes produced is greater in the case of powders. When the powder is irradiated by laser pulses, the diffusion rate of C$_2$ radicals into the solvent is larger and the density lower compared to the pellet, in which radicals are formed locally on its surface in high density promoting longer polyynes. However, powder samples has a greater surface area compared to that of the pellets, which allows the formation of a larger number of C$_2$ radical, hence increasing the total amount of polyynes.
Hydrogen-capped polyynes up to 14 carbons were obtained ablating graphite powder in the work of Wakabayashi \textit{et al.} \cite{waka_2012} and up to 16 carbons in the work of Tabata \textit{et al.} \cite{tabata2006}.
\\Tsuji \textit{et al.} employed fullerene particles obtaining hydrogen-capped polyynes up to 12 carbon atoms \cite{tsuji2003}. C$_2$ radicals coming from C$_{60}$ particles undergo polymerization and hydrogenation of carbon clusters.

\paragraph{Direct ablation of the solvent}\
\\Polyynes were produced, even when no target was used, simply by irradiating pure solvents with femtosecond laser pulses. In particular, the employed solvents are acetone \cite{hu2008}, toluene \cite{rama2015, rama2017}, n-hexane \cite{sato2010}, benzene \cite{weso2011}, and n-octane \cite{zaidi2010,zaidi2019femtosecond,sato2010}. Indeed, the power densities reached in the focus of femtosecond lasers, up to 10$^{15}$ W/cm$^2$, are so high that can cause the direct dissociation of the solvent itself in atomic carbon and carbon radicals, allowing the growth of polyynes without the need for a physical target \cite{hu2008,rama2017,sato2010,weso2011,zaidi2010}. All the works listed in the "No Target" section of \cref{table:target} agree with a common description of the formation mechanism of polyynes by \textit{fs} laser pulses. As described in \cref{formationmec_sec}, short polyynes as HC$_2$H or HC$_4$H can be produced directly from the dissociation of the solvents as hydrocarbons and alkanes \cite{sato2010,rama2017,weso2011,zaidi2010}. Long polyynes (\textit{i.e.} longer than 6 carbon atoms) seemed not to be formed in a single-step way by parent ions but are the product of secondary reactions of fragments and radicals formed during the dissociation of the solvent by the action of the \textit{fs}-laser.
\\In particular, ablation in pure acetone by Hu \textit{et al.} allowed the synthesis of HC$_6$H that has been detected by surface-enhanced Raman spectroscopy \cite{hu2008}. The power density employed in this case was estimated to be approximately 10$^{15}$ W/cm$^2$.
\\Sato \textit{et al.} synthesized hydrogen-capped polyynes up to 12 carbon atoms from ablations in n-hexane and n-octane \cite{sato2010}. They found that by employing n-octane (C$_{10}$H$_{22}$) as solvent there is no evident enrichment of the concentration of HC$_{10}$H as expected, confirming that direct parent ions do not lead to the formation of long polyynes \cite{sato2010}. Ablation in n-octane was carried out also by Zaidi \textit{et al.} who found hydrogen-capped polyynes up to HC$_{14}$H \cite{zaidi2010,zaidi2019femtosecond}.
\\Benzene and its deuterated form, instead, were employed by Wesolowski \textit{et al.} with the aim of producing both polyynes and amorphous carbon nanoparticles \cite{weso2011}. They synthesized hydrogen-capped polyyne up to 14 carbon atoms. Moreover, they found that linear or cyclic polyyne may act as bridges between sp$^2$-carbon aromatic rings, building up a peculiar sp-sp$^2$ amorphous carbon material.
\\Furthermore, toluene was recently employed as a solvent for \textit{fs}-laser ablation and demonstrated to be an efficient solvent for the formation of long polyynes \cite{rama2017}. Indeed, hydrogen-capped polyynes were detected ranging from 12 to 18 carbon atoms, while shorter chains were produced but not observed by the chosen HPLC method. Moreover, methyl-capped polyynes (HC$_n$CH$_3$) with 12 and 14 carbon atoms were also identified \cite{rama2017}. The formation of methyl-capped polyynes was due to the dissociation of toluene that can release several ions with methyl groups inside. Once again, the production of long polyynes cannot come from single-step rearrangement of parent ions of toluene but is due to secondary reactions \cite{rama2017}. Moreover, cumulenes can also be synthesized but they decay too quickly to be detected \cite{rama2017}.

\section{Sp-carbon chains-based nanocomposites}
In this Section, we focus on works in which polyynes are embedded in polymer matrices marking a first step for the potential use of these nanocomposite for new technological applications. There are three approaches to prepare sp-carbon chains-based nanocomposites. The first method involves the production of polyynes by PLAL, which were then embedded in a polymeric solution of PVA, left to dry to obtain the nanocomposite film \cite{okada2011}. The second one is based on the dipping of a PVA film in a solution of boric acid and polyynes by PLAL \cite{sata2019}. The last approach is the method developed in our laboratory in which polyynes were directly produced in a polymeric solution, where PLAL was performed \cite{peggianiPVA, tesiSacchi, tesiPeggiani}.
Polymers are characterized by a large variety of Young modulus, tensile strength, electrical and thermal conductivity. Moreover, they confer lightweight and flexibility when used as a matrix in nanocomposites at a low cost.
The few works in the literature dealing with the fabrication of nanocomposites with polyynes synthesized by pulsed laser ablation in liquid are reported in \cref{table:nanocomposite}. 

\begingroup
  \footnotesize 
  \centering

    \begin{longtable}{@{\extracolsep{\fill}}|l@{ } |l@{ }|p{3.1 cm}|l@{ }|l@{ }|l@{ }|l@{ }|l@{ }|@{}} 
          \caption{Polyynes by PLAL embedded in polymeric matrices.\label{table:nanocomposite}}\\ 
\hline
    \textbf{Matrix}
    &\textbf{Solvent}
    &\textbf{Target}    
    & \boldsymbol{$\lambda$} 
    & \begin{tabular}[c]{@{}l@{}}\textbf{Frequency},\\ \textbf{Duration}\end{tabular}
    & \boldsymbol{$\tau$}                                        
    & \begin{tabular}[c]{@{}l@{}}\textbf{Fluence, Energy Pulse},\\ \textbf{Spot Size}\end{tabular}
    & \textbf{Polyynes }
    \\ \hline

PVA \cite{okada2011}
                        & methanol
                        & graphite powder
                        & 532 nm
                        & 20 Hz
                        & 7 ns
                        & 1.2 J/cm$^2$
                        & -
    \\\hline
PVA \cite{sata2019}
                        & methanol$^a$ 
                        & carbon particles
                        & -
                        & -
                        & -
                        & -
                        & C$_n$H$_2$ (n=10-14)
    \\\hline
PVA  \cite{peggianiPVA}  
                        & water
                        & graphite pellet
                        & 532 nm
                        & 10 Hz, 15-30 min
                        & 5-7 ns
                        & 5.2 J/cm$^2$, 1.4 mm
                        & C$_n$H$_2$ (n=6-10)  
     \\\hline
PVA \cite{tesiSacchi}
                        &\begin{tabular}[t]{@{}l@{}} water:acetonitrile;\\ water:methanol;\\ water:ethanol \end{tabular}
                        &\begin{tabular}[t]{@{}l@{}}Ag pellet;\\Ag pellet + graphite pellet\end{tabular}
                        & 532 nm
                        &\begin{tabular}[t]{@{}l@{}}10 Hz, 15 min;\\ 10 Hz, 15 min + 15 min\end{tabular}
                        & -
                        &\begin{tabular}[t]{@{}l@{}} 10.06 J/cm$^2$, 30 mJ;\\10.15 J/cm$^2$, 30 mJ;\\9.93 J/cm$^2$, 30 mJ\end{tabular}
                        & -
    \\\hline 
PMMA  \cite{tesiPeggiani}  
                        & acetone
                        & Ag pellet + graphite pellet
                        & 532 nm
                        & 10 Hz, 5 min + 15 min
                        & 5-7 ns
                        & 2 J/cm$^2$
                        & -
    \\\hline

\multicolumn{8}{l} {\footnotesize{$^a$ Boric acid is mixed with the size-selected polyynes and then PVA is added to the solution}}\\

  \end{longtable}
  
\endgroup

\begin{figure}[!ht] 
	\centering
	\hspace{8mm}
	\subfloat[][]{\includegraphics[scale=0.35]{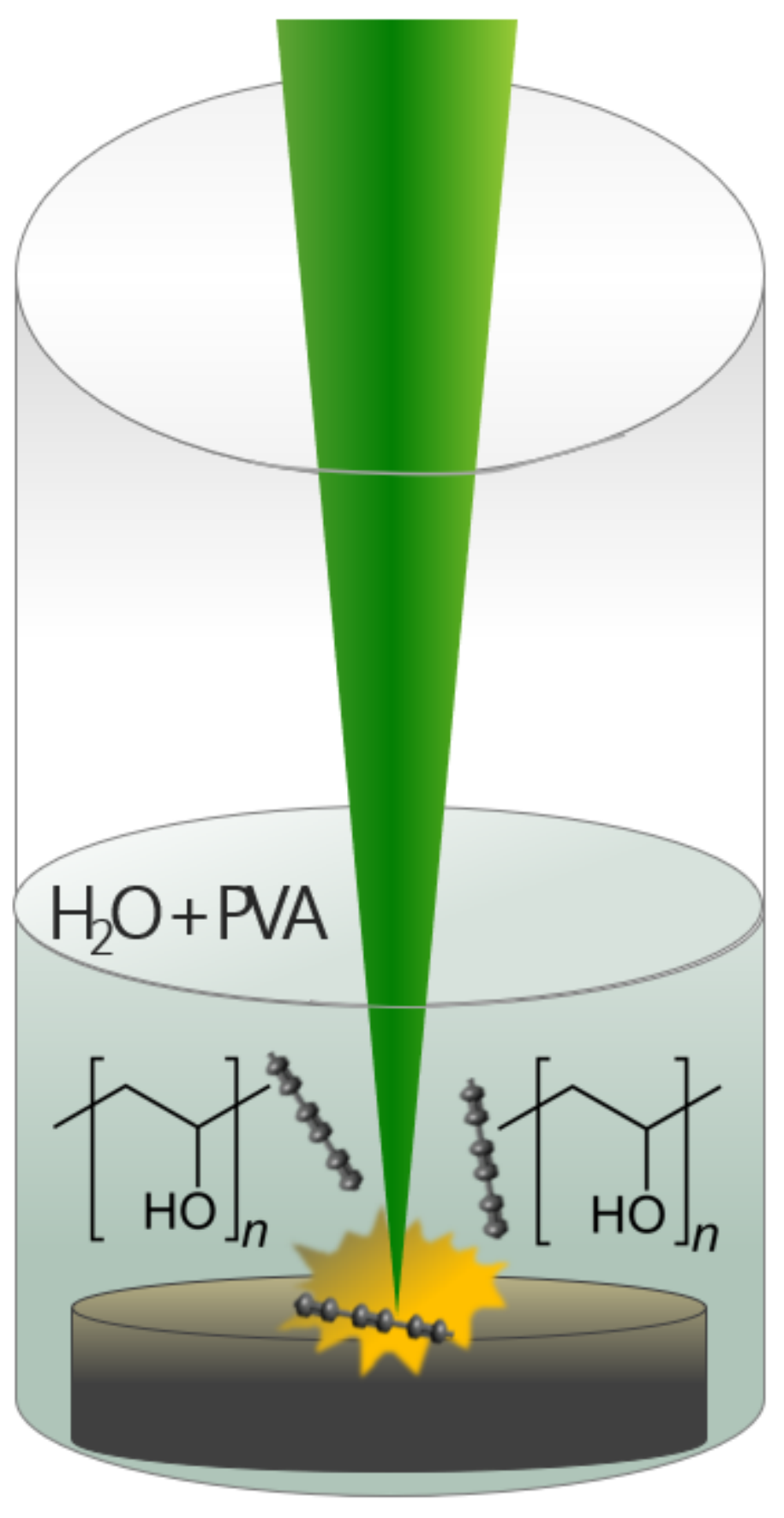}\label{fig:pva_insitu}}
	\hspace{14mm} 
	\subfloat[][]{\includegraphics[scale=0.25]{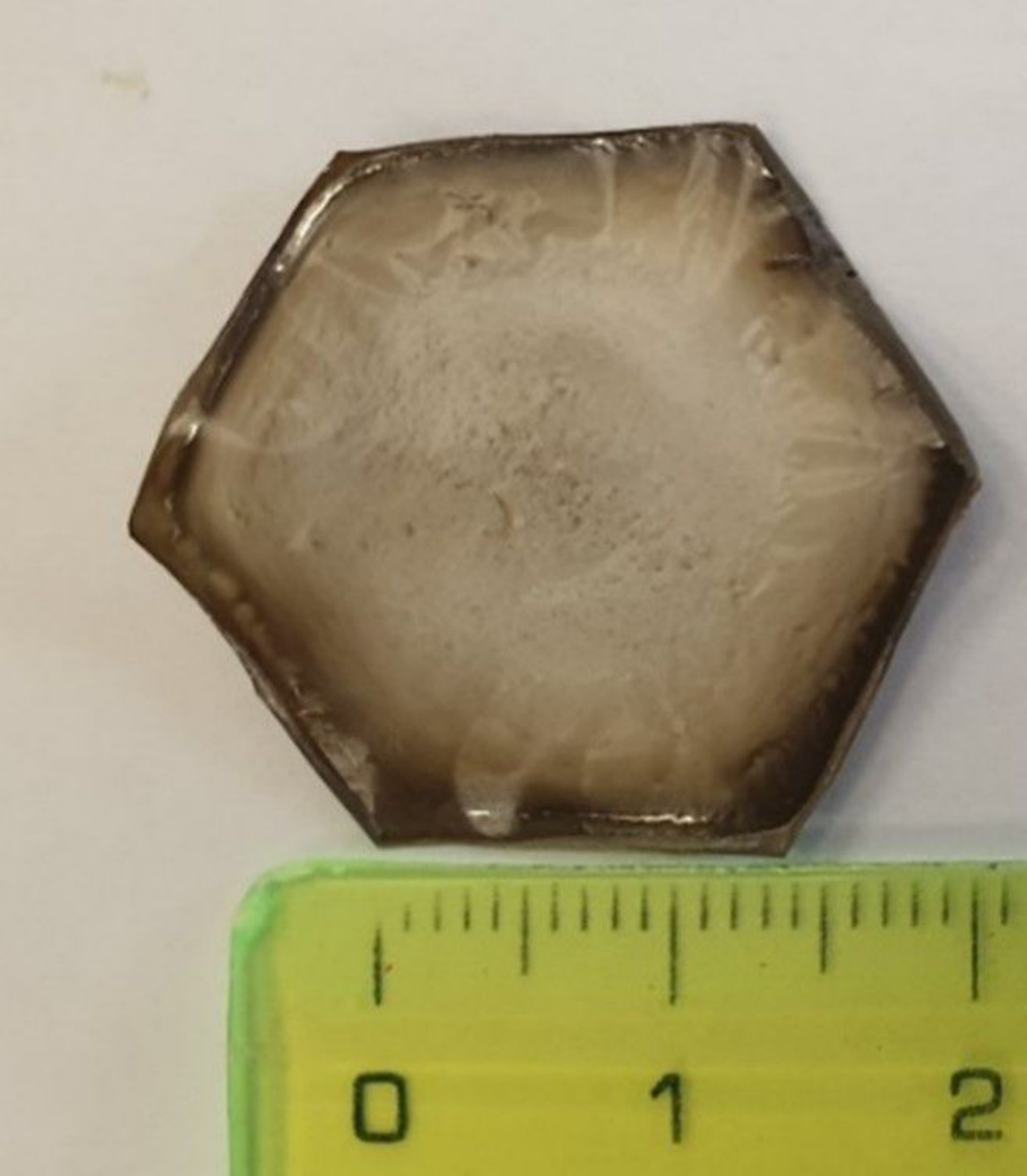}\label{fig:sample_pva}}
	\hspace{2mm} 
	\subfloat[][]{\includegraphics[scale=0.47]{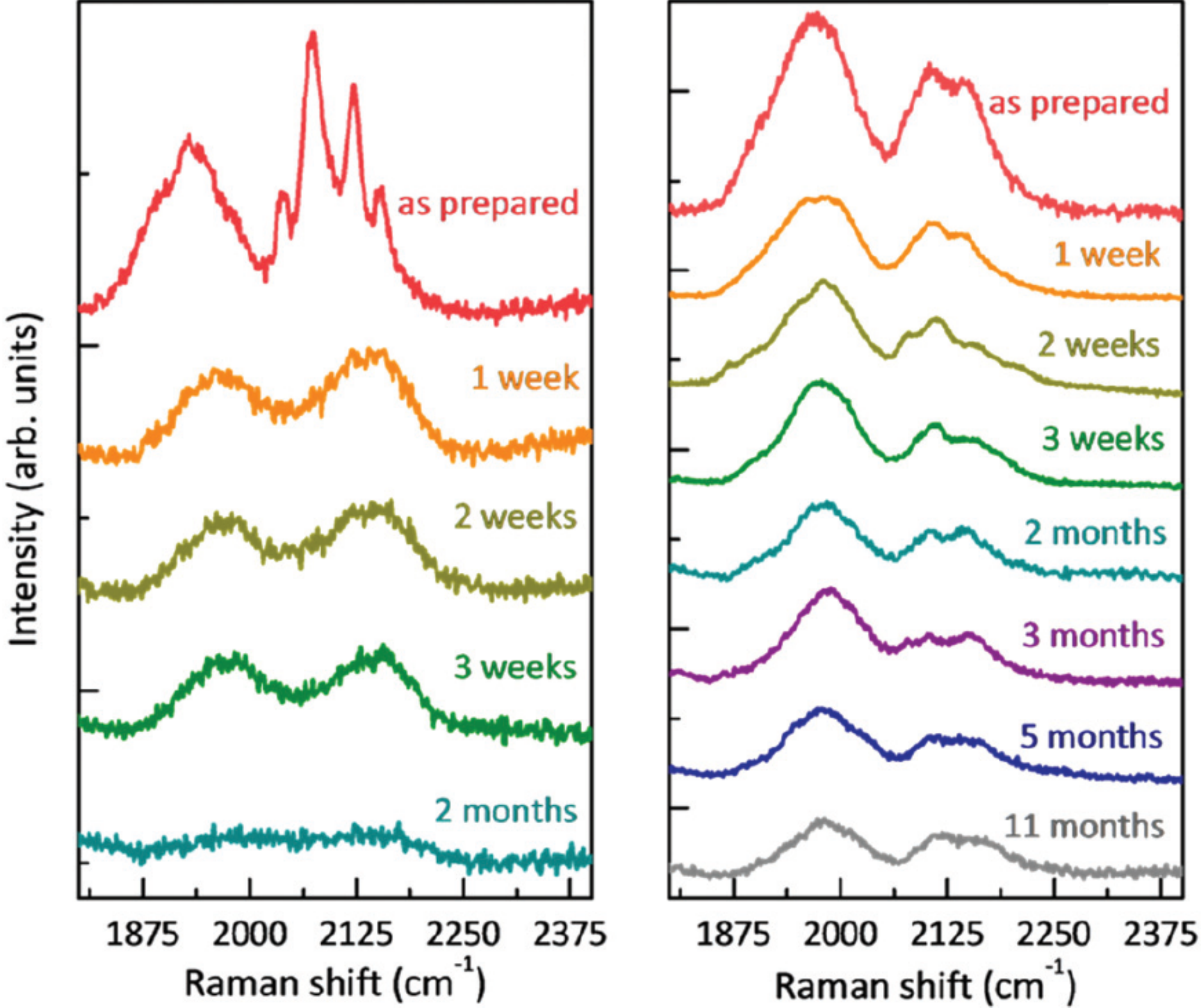}\label{fig:PVA_polyyne}}
	\caption{\protect\subref{fig:pva_insitu} Sketch of the pulsed laser ablation in an aqueous solution of PVA. \protect\subref{fig:sample_pva} Free-standing film based on PVA, Ag nanoparticles and polyynes. \protect\subref{fig:PVA_polyyne} Evolution in time of SERS spectra (excitation line at 514.5\,nm): (left) Ag/PVA/polyynes liquid sample and (right) Ag/PVA/polyynes nanocomposite.} 	
\end{figure}

As shown in Table\ref{table:nanocomposite} the majority of the works employed poly(vinyl alcohol) (PVA) as a matrix due to several advantageous properties of this polymer. Indeed, PVA is widely exploited in the formation of carbon or metal-based composites because it is a low-cost, chemically stable material, and soluble in water displaying also good film formation properties \cite{mallakpoura2017,pandey2011}.\\ 
Okada \textit{et al.} \cite{okada2011} were the first group, in 2011, to fabricate a nanocomposite film by employing PVA to prevent the tendency of polyynes to transform into more stable sp$^2$ structures.
They ablated a suspension of graphite powder dispersed in methanol under the action of a magnetic stirrer to homogenize the solution and favor the formation of polyynes. The polyynes solution was then mixed with Ag colloidal solution in a volume ratio of 1:4. For the fabrication of the nanocomposite film they added PVA granules to the polyynes/Ag solution employing a temperature of 373 K to allow the complete dissolution of the polymer. Then they casted the solution on glass substrate and uniformly dried in air for 48 h at 303 K to obtain the nanocomposite film. In order to assess the stability of polyynes in nanocomoposite film SERS spectra were measured in time proving that the sp-carbon chains were stable for at least a month. UV-Vis spectra were also carried out both on PVA/Ag colloid and on the nanocomposite film. For what concerns the PVA/Ag colloid a narrow absorption band located at 425 nm corresponding to the surface plasmon resonance (SPR) of Ag nanoparticles was detected. The result on the nanocomposite film showed instead that the SPR absorption band was shifted to a longer wavelength and broadened. This feature was already attributed to aggregation of Ag nanoparticles when mixed with polyyne solutions \cite{Urso2010}.\\
More recently Sata \textit{et al.} \cite{sata2019} investigated angular dependence of absorption intensity of linearly polarized UV light in solid films embedding aligned polyyne molecules. This work goes in the direction of real application since the study of linear dichroism (LD) is of great importance for liquid crystal display (LCD) where polarized absorption of light is utilized under the control of molecular alignments \cite{Norden}.
The research group started from the fabrication of PVA film after drying a solution of PVA distilled water in a flat container of stainless steel for a week. Laser ablation of carbon particles in methanol was performed and followed by separation using HPLC eluted by methanol in order to size-select three polyynes(i.e. C$_n$H$_2$, n=10-14). 
Then a PVA film was dipped in the polyyne/boric acid (H$_3$BO$_3$) solution in methanol at 313 K for 3 h. Boric acid was added to the solution to harden the nanocomposite film by favoring the crosslinking bonds between the polymer chains. In order to align the polyynes trapped in the interstitial space of the polymer, the nanocomposite film was stretched in one direction and then held in place at the desired length and allowed to dry to recover its hardness.
To verify the alignment of the sp-carbon chains inside the PVA film the authors employed a home-built low-magnitude UV microscope equipped with a light source of linearly polarized light. The performed measurements revealed that the absorption
intensity was maximum for each vibronic band of the size-selected polyynes when the polarization angle of the incident light was in parallel with the stretching direction of the nanocomposite film. Moreover, the intensity of the maximum absorption peak of the chains decreases reaching the minimum when the incident light is perpendicular to the stretching direction of the nanocomposite film.\\
For the first time, we synthesized polyynes by performing the laser ablation in a solution in which PVA was already dissolved \cite{peggianiPVA}. The choice to employ the polymer directly in the ablation solution (see Figure \ref{fig:pva_insitu}) can be justified by the following reasons. When the polymer is added after the ablation process, the solution containing the synthesized polyynes needs to be heated to allow the dissolution of the polymer affecting the sp-carbon chains integrity since they are susceptible to temperature rises \cite{casari2004}. Another aspect related to the addition of the polymer after the synthesis of polyynes is that it could prevent the complete blending between polyynes, metal nanoparticles and polymer. Indeed, to overcome these critical issues in the preparation of polyynes-based nanocomposite, we performed the ablation of a graphite solid pellet in an aqueous solution of PVA obtaining hydrogen-capped polyynes up to 10 carbon atoms. We also investigated the role of the polymer concentration affecting the viscosity of the solution finding an optimum value, both for the ablation process and for the nanocomposite fabrication, of 1 wt.\%.
In order to perform SERS analyses on the nanocomposite film, Ag nanoparticles were prepared using the Lee–Meisel method \cite{npLee}, and then added to the PVA/polyynes solution. To fabricate the nanocomposite film (see Figure \ref{fig:sample_pva}), the Ag/PVA/polyynes solution was dropcasted in a small plastic vessel and left to dry for 24 hours at room temperature. To asses if there was an improvement of the stability of the sp-carbon chains we performed SERS measurements over time on both Ag/PVA/polyynes liquid and solid samples, see \cref{fig:PVA_polyyne}. 
\\In the case of the liquid, after just 1 week the well-defined polyyne features of the SERS spectrum were replaced by two broad bands, still detectable after 3 weeks, and after 2 months these characteristic features were no longer visible. For what concerns the nanocomposite, we observed a decrease in the intensity of the polyynic SERS bands after 1 week but no other substantial changes occurred in the following 11 months. Our results indicate that the stability of polyynes, especially in the case of  nanocomposite, is greatly enhanced due to the presence of the solid matrix which inhibits the reactions between sp-carbon chains by immobilizing them. As a final consideration this simple method of synthesis of hydrogen-capped
polyynes in PVA was performed in water, which is a low-cost and environmental friendly solvent, proving to be a promising route for future optical and mechanical characterization of polyyne-based films.\\
We also performed PLAL in solution where PVA was dissolved in a mixture of water and organic solvents \cite{tesiSacchi}. In this work, PVA was dissolved in water and then three different solutions were prepared by adding three organic solvents: acetonitrile, ethanol and methanol. The addition of organic solvents was investigated in order to increase polyynes production during the ablation process. For what concerns the PVA/acetonitrile solution it was possible to reach up to 42\% vol of ACN without having PVA aggregation phenomena. The ablation was the performed using an Ag pellet as a target in order to obtain in one step both polyynes and Ag nanoparticles for the subsequent SERS characterization. A nanocomposite film was fabricated by drop-casting technique on Si substrate and the SERS analysis revealed that the polyynes signals were visible at least up to 18 days. In the case of methanol and ethanol a double ablation process was performed since these solvents have a lower yield of polyynes with respect to acetonitrile \cite{peggianisolventi}. The two-step ablation process involved a 15 minutes  ablation of Ag pellet followed by 15 minutes ablation of graphite pellet. All the ablations with organic solvents were performed in 1.5 ml polymeric solution, with a laser energy of 30 mJ. For what concerns methanol the volume fraction added to PVA/water solution was up to 75\%, instead with ethanol it was possible to reach up to 50\%. The nanocomposite film obtained from these solutions showed a SERS  signals in the polyynes region at least up to 6 weeks and 8 weeks for ethanol and methanol, respectively.\\
We were the first to investigate the synthesis of polyynes by pulsed laser ablation in a PMMA solution \cite{tesiPeggiani}. 
PMMA was selected because it is transparent, high-compatible with silver nanoparticles and insoluble in water. This latter aspect allows the preparation of a water-resistant nanocomposite, which is of interest for waterproof technological applications. Also in this work a two-step ablation was performed. The first step ablation of Ag pellet was carried out in a PMMA/acetone solution for 5 minutes. The second step involved the ablation of a graphite pellet for 15 minutes in the solution obtained after the first step. 
Although the concentration of Ag nanoparticles was several orders of magnitude lower than those obtained by chemical synthesis the SERS signal was clearly visible once the nanocomposite film was fabricated. The nanocompostie film was monitored in time showing that the two main bands of carbon-atom wires were preserved at least for 5 months. The result of our work demonstrates that a prolonged stability, compared to hydrogen-capped polyynes not encapsulated in a polymeric matrix, was achieved.


\section{Conclusions}
We have here reviewed the state of the art of research on the synthesis of polyynes by PLAL. PLAL is a versatile technique able to efficiently produce different types of polyynes with various terminations and lengths. The use of laser ablation enables the adoption of different targets and the \textit{contactless} process based on laser-matter interaction allows the adoption of \textit{in situ} characterization techniques aimed at understanding the formation mechanisms. The solvent turns out to play a fundamental role in the production yield and in providing the chain terminations. Notwithstanding the recent progress, work still needs to be done to shed light on the complex mechanisms leading to the formation of polyynes. Such mechanisms include laser-target interaction, target ablation, plasma formation, and expansion in a liquid environment, polyyne formation and diffusion in the liquid. Some works have shown the use of \textit{in situ} characterization techniques to get some information into the fundamental processes at play, even though much efforts still need to be spent to fully illustrate the formation mechanisms of sp-carbon chains. The possibility to produce different chains with tailored terminations and the realization of solid materials with long-term stability can open to the exploitation of the peculiar properties of sp-carbon chains in advanced applications. The peculiar nonlinear optical properties are appealing in optoelectronics while applications in biosensing and organic electronics have been recently reported \cite{marabotti2021vibrational,pecorario2022stable,scaccabarozzi2020field}.

\section*{Acknowledgements}
The authors acknowledge funding from the European Research Council (ERC) under the European Union's Horizon 2020 research and innovation program ERC Consolidator Grant (ERC CoG2016 EspLORE grant agreement no. 724610, website: www.esplore.polimi.it). We thank A. Sacchi and M. Benoliel for their assitance during the measurements and the data analyses.

\bibliography{references} 
\bibliographystyle{iopart-num-mod}

\end{document}